%% file: main.tex
\documentclass[sigconf]{acmart}

\usepackage{amsmath}
\usepackage{algorithmic}
\usepackage{graphicx}
\usepackage{textcomp}
\usepackage{xcolor}
\usepackage{enumitem}
\usepackage{tabularx}
\usepackage{booktabs}
\usepackage{multirow}
\usepackage{verbatim}
\usepackage{subcaption}

\setlist[itemize]{leftmargin=*}
\setlist[enumerate]{leftmargin=*}

\usepackage{algorithm}
\usepackage[most]{tcolorbox}
\definecolor{titleblockcolor}{HTML}{353535}
\definecolor{textblockcolor}{HTML}{FFFFFF}
\newenvironment{block}[2][]{
  \begin{tcolorbox}[adjusted title=#2, fonttitle={\small\bfseries}, colback={textblockcolor}, colframe={titleblockcolor}, coltitle={white}, arc=0pt,
  outer arc=0pt, left=1pt, right=1pt, fontupper=\small, #1, breakable]
}{\end{tcolorbox}}

\newtcbox{\code}{
  on line,
  boxsep=0pt, left=3pt, right=3pt, top=2pt, bottom=2pt,
  colback=gray!20,
  colframe=gray!20,
  arc=3pt,
  boxrule=0pt,
  fontupper=\ttfamily
}

\AtBeginDocument{%
  }

\setcopyright{acmlicensed}
\copyrightyear{2018}
\acmYear{2018}
\acmDOI{XXXXXXX.XXXXXXX}
\acmConference[Conference acronym 'XX]{Make sure to enter the correct
  conference title from your rights confirmation email}{June 03--05,
  2018}{Woodstock, NY}

\begin{document}

%%
%% Title
%%
\title{CacheRAG: A Semantic Caching System for Retrieval-Augmented Generation in Knowledge Graph Question Answering}

%%
%% Authors
%%
\author{Yushi Sun}
\orcid{0000-0003-3853-6364}
\email{ysunbp@connect.ust.hk}
\affiliation{%
  \institution{HKUST}
  \city{Hong Kong}
  \country{China}
}

\author{Lei Chen}
\orcid{0000-0002-8257-5806}
\email{leichen@cse.ust.hk}
\affiliation{%
  \institution{HKUST(GZ) / HKUST}
  \city{Guangzhou / Hong Kong}
  \country{China}
}

\renewcommand{\shortauthors}{Sun and Chen}

%%
%% Abstract
%%
\begin{abstract}
The integration of Large Language Models (LLMs) with Retrieval-Augmented Generation (RAG) has significantly advanced Knowledge Graph Question Answering (KGQA). However, existing LLM-driven KGQA systems act as stateless planners, generating retrieval plans in isolation without exploiting historical query patterns: analogous to a database system that optimizes every query from scratch without a plan cache. This fundamental design flaw leads to schema hallucinations and limited retrieval coverage. We propose CacheRAG, a systematic cache-augmented architecture for LLM-based KGQA that transforms stateless planners into continual learners. Unlike traditional database plan caching (which optimizes for frequency), CacheRAG introduces three novel design principles tailored for LLM contexts: (1) Schema-agnostic user interface: A two-stage semantic parsing framework via Intermediate Semantic Representation (ISR) enables non-expert users to interact purely in natural language, while a Backend Adapter grounds the LLM with local schema context to compile executable physical queries safely. (2) Diversity-optimized cache retrieval: A two-layer hierarchical index (Domain $\rightarrow$ Aspect) coupled with Maximal Marginal Relevance (MMR) maximizes structural variety in cached examples, effectively mitigating reasoning homogeneity. (3) Bounded heuristic expansion: Deterministic depth and breadth subgraph operators with strict complexity guarantees significantly enhance retrieval recall without risking unbounded API execution. Extensive experiments on multiple benchmarks demonstrate that CacheRAG significantly outperforms state-of-the-art baselines (e.g., +13.2\% accuracy and +17.5\% truthfulness on the CRAG dataset).
\end{abstract}

%%
%% CCS Concepts
%%


%%
%% Keywords
%%

\maketitle

\input{secs/sec_intro}

\input{secs/sec_related}

\input{secs/sec_problem}

\input{secs/sec_method}
\input{secs/sec_experiment}
\input{secs/sec_conclusion}

%%
%% Acknowledgments
%%
%\begin{acks}
%This work was partially supported by ...
%\end{acks}

\bibliographystyle{ACM-Reference-Format}
\bibliography{references}
\appendix
\input{secs/appendix}

\end{document}

%% file: secs/sec_intro.tex
\section{Introduction}
\label{sec:intro}
Knowledge Graph Question Answering (KGQA) is a fundamental task in data management and artificial intelligence, aiming to provide precise answers to natural language questions over structured knowledge bases. Recently, the integration of Large Language Models (LLMs) via the Retrieval-Augmented Generation (RAG) paradigm has emerged as a promising solution. By incorporating semantic parsing and query plan generation into the LLM's reasoning process, these RAG-based systems can handle complex queries and generate context-aware responses.

\begin{figure*}[hbtp]
  \centering
  \includegraphics[width=0.9\linewidth]{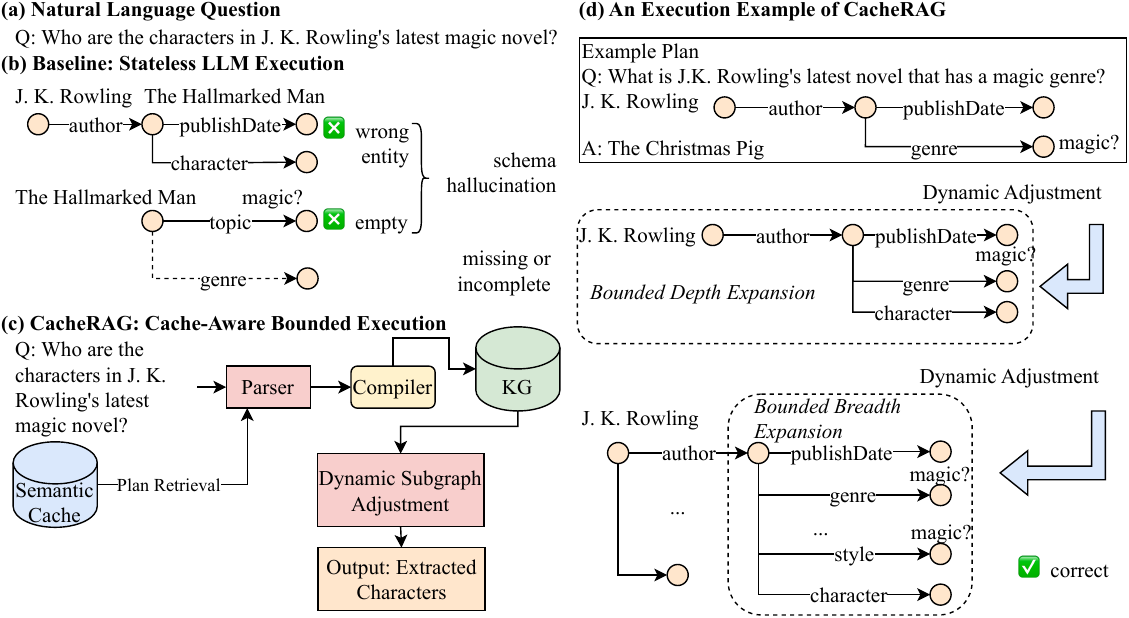}
  \vspace{-1em}
  \caption{Comparison of stateless LLM execution (baseline) and CacheRAG (our approach) on a KGQA task. (a) Input: natural language question about J.K. Rowling's latest magic novel. (b) Baseline suffers from schema hallucination (retrieving wrong entity \textit{The Hallmarked Man}, retrieval of non-existent attributes such as \code{topic}) and incomplete retrieval (missing the \code{genre} attribute). (c) CacheRAG workflow: the parser extracts ISR, semantic cache provides historical plans via plan retrieval, and bounded subgraph expansion dynamically adjusts the query plan. (d) Execution example: CacheRAG refers to the useful cached example, applies bounded depth/breadth expansion with dynamic adjustment to explore alternative schema paths (\code{genre}, \code{style}, \code{publishDate}, \code{character}), successfully identifying the correct KG subgraph for QA.}
  \label{fig:teaser}
  \vspace{-1em}
\end{figure*}

However, when deployed in real-world scenarios, existing LLM-driven KGQA pipelines expose critical system-level flaws. Most prominently, these systems act as \textbf{stateless planners}: they generate retrieval plans (e.g., SPARQL queries or API calls) based solely on the current question, \textbf{failing to leverage historical question-answering experiences}. This design mirrors a hypothetical database system that optimizes every query from scratch without a plan cache: an inefficiency that modern DBMSs resolved decades ago through execution plan caching~\cite{graefe1993volcano}. Yet, naively adapting traditional DB plan caching to LLM-based systems fails due to fundamental differences: (1) LLM context windows reward \textit{diversity} over frequency (unlike LRU/LFU), and (2) retrieval plans must be \textit{schema-aligned} to prevent hallucinations.

The stateless KGQA reasoning approach inevitably leads to two major problems, as illustrated in Figure~\ref{fig:teaser} (b):
First, \textbf{susceptibility to schema hallucination}. Consider the query ``Who are the characters in J. K. Rowling's latest magic novel?'' Existing stateless LLM planners might blindly retrieve her latest general novel (\textit{The Hallmarked Man}, a crime fiction) by missing the ``magic'' constraint. When forced to self-correct, they frequently hallucinate schema attributes, attempting to query a non-existent \code{topic: magic} predicate, leading to empty results and execution failures.

Second, \textbf{limited retrieval coverage}. Even when the entity is correctly identified, zero-shot planning often generates shallow, incomplete query structures: missing multi-hop reasoning paths or failing to explore alternative schema branches when initial retrieval returns empty results.

\paragraph{Root Cause and Our Solution.} 

\textbf{Root Cause 1: Schema Misalignment}. Existing systems expect users to phrase queries using exact KG predicates (e.g., ``novels with topic=magic'') instead of natural language. Our CacheRAG method addresses this through a two-stage decoupling architecture:
\begin{enumerate}[leftmargin=*,label=(\arabic*)]
\item \textbf{User Layer (Logical)}: Users express intents naturally (``magic novel''). The LLM acts as a syntactic parser to extract an Intermediate Semantic Representation (ISR): \{entity: J.K. Rowling, type: novel, constraints: [magic]\}.
\item \textbf{System Layer (Physical)}: The Backend Adapter grounds the LLM by fetching the actual local schema (e.g., valid edges around the retrieved entity) and prompts the LLM to compile the logical ISR into an executable physical query. For instance, given the constraint ``magic'' and a local schema containing \code{genre}, the LLM safely maps the intent to the \code{genre: fantasy} predicate. This eliminates the need for user queries to match exact domain-specific predicate names while substantially mitigating schema hallucination.
\end{enumerate}

\textbf{Root Cause 2: Lack of Historical Query Patterns}. Traditional stateless planners reinvent query logic for every question, wasting API calls and producing shallow plans. As shown in Figure~\ref{fig:teaser}(c), CacheRAG introduces a Semantic Cache that stores successful historical retrieval plans. During plan generation, the cache retrieval module provides useful examples (e.g., similar queries demonstrating ``author $\rightarrow$ novels $\rightarrow$ genre'' navigation patterns), enabling the LLM to learn complex multi-hop reasoning rather than relying on zero-shot guessing. When initial retrieval is insufficient, the heuristic dispatcher (Figure~\ref{fig:teaser}(d)) triggers bounded depth/breadth expansion, exploring alternative schema paths (e.g., checking both \code{genre} and \code{style} predicates) until the correct answer is found.

This dual-stage design ensures that non-expert users interact purely via natural language (addressing Root Cause 1), while the system leverages historical query patterns to guide plan generation (addressing Root Cause 2): eliminating the need for users to learn KG-specific predicates or SPARQL syntax (Section 3.1).

\textbf{Our Key Insight.} 
Successful KGQA interactions exhibit strong \textit{structural reusability}: for example, questions about movie awards consistently require similar multi-hop patterns (e.g., film $\rightarrow$ director $\rightarrow$ other films $\rightarrow$ nominations), regardless of specific entities. Yet, stateless LLMs must rediscover these patterns for every query. CacheRAG bridges this gap by introducing a systematic cache-augmented architecture for LLM-based KGQA, transforming stateless planners into continual learners that exploit historical query patterns while maintaining rigorous schema fidelity and execution bounds. Designing such a robust cache-augmented system requires systematically addressing four fundamental challenges. For each challenge, CacheRAG introduces a targeted architectural design choice with strict justifications:

\begin{itemize}   
\item \textbf{Challenge 1: Non-Deterministic Attribute Extraction.} LLMs often hallucinate predicates (e.g., extracting non-existent attribute ``topic''), leading to execution failures and making the system unintuitive for non-expert users.

\textbf{Solution 1: Schema-Constrained Semantic Parsing.} We decouple the user-facing prompt from KG-specific jargon. The LLM first extracts an Intermediate Semantic Representation (ISR). A Backend Adapter then provides the valid local schema context, repurposing the LLM to compile the ISR into physical KG queries. This ensures schema fidelity and enables \textbf{non-expert users to interact purely via natural language} without learning SPARQL or KG schemas.

\item \textbf{Challenge 2: Suboptimal Experience Replay (Cache Scheduling).} Traditional cache metrics (LRU/LFU) or naive semantic similarity fall short in LLM contexts because injecting structurally identical cached plans wastes the LLM's limited token window and induces mode collapse, providing minimal information gain.

\textbf{Solution 2: Diversity-Aware Cache Management.} We enhance cache retrieval by utilizing a two-layer hierarchical index (Domain $\rightarrow$ Aspect) and a Maximal Marginal Relevance (MMR) scoring function. This guarantees the LLM is provided with a structurally diverse set of topological query patterns, maximizing its multi-hop reasoning capabilities.

\item \textbf{Challenge 3: Limited Retrieval Coverage in Complex QA.} LLMs often fail to retrieve sufficient context in a single pass, yet unbounded retry loops risk exhaustive API calls and system crashes.

\textbf{Solution 3: Dynamic Plan Adjustment via Bounded Expansion.} We introduce dynamic plan adjustment governed by an auto-termination judge. If initial retrieval is insufficient, the system triggers strictly bounded execution operators: Depth Expansion (auto-chaining) and Breadth Expansion (star queries). This significantly enhances retrieval recall while maintaining strict algorithmic bounds on execution complexity.

\item \textbf{Challenge 4: The Cold-Start Dilemma.} At system initialization, the cache is empty, leaving the LLM with no reference plans to navigate complex schemas.

\textbf{Solution 4: Offline Auto-Generation.} We introduce an exploratory view materialization strategy. By sampling the KG schema offline to synthesize initial $\langle \text{Query}, \text{Plan}, \text{Answer} \rangle$ tuples, CacheRAG ensures high-quality continual learning from the very first user query.

\end{itemize}

In summary, \textbf{this paper makes the following contributions}:

\begin{itemize}
    \item \textbf{Cache-Augmented KGQA System.} We propose CacheRAG, a stateful cache-augmented architecture that transforms stateless LLM-based KGQA planners into history-aware reasoning engines. To the best of our knowledge, CacheRAG is the \textbf{first system to integrate semantic plan caching specifically designed for LLM continual learning} in KGQA: distinct from infrastructure-level KV-caches~\cite{agarwal2025cache,jin2025ragcache} or general thought-template buffers~\cite{yang2024buffer}, which neither target schema-constrained KG traversal nor optimize for structural diversity in in-context learning.

    \item \textbf{Schema-Agnostic Semantic Parsing via ISR.} We design a two-stage ISR parsing framework with a Backend Adapter that \textbf{achieves 98.9\% physical query compilation success} and 98\% entity extraction accuracy on CRAG, substantially mitigating schema hallucination without requiring users to learn KG predicates or SPARQL. The two-stage logical/physical decoupling is analogous to query compilation in traditional DBMSs, adapted here to the heterogeneous schema environment of LLM-driven KG retrieval.

    \item \textbf{Diversity-Aware Cache Retrieval.} We adapt MMR-based retrieval to a two-layer hierarchical index (Domain~$\rightarrow$~Aspect), enforcing structural diversity within the LLM's limited context window. Compared to static few-shot exemplars and naive BM25 retrieval, this design \textbf{improves accuracy by 4.3\% and truthfulness by 5.5\%} (ablation, Section~\ref{subsec_ablation}), preventing redundancy and reasoning mode collapse.

    \item \textbf{Bounded Graph Expansion Operators.} We formalize depth ($\sigma_{depth}$, Index Nested Loop Join, $K_{depth} \le 3$) and breadth ($\sigma_{breadth}$, Star-Pattern Scan, $\mathcal{O}(K_{depth} \cdot K_{degree})$ space) expansion operators with \textbf{strict complexity guarantees}, improving retrieval recall from 0.756 to 0.927 while preventing unbounded API calls and OOM failures. Controlled ablation confirms cache-guided expansion outperforms blind expansion by up to 6\%.

    \item \textbf{Extensive Empirical Validation.} Experiments on four benchmarks (CRAG, QALD-10-en, WebQSP, CWQ) show CacheRAG outperforms state-of-the-art by \textbf{+13.2\% accuracy and +17.5\% truthfulness} on CRAG, and achieves new SOTA on all three SPARQL-based benchmarks, without any fine-tuning.
\end{itemize}

%% file: secs/sec_related.tex
\section{Related Work}
\label{sec_related}
KGQA approaches can be broadly classified into two categories: Semantic Parsing KGQA and KG-based RAG. Additionally, we contextualize our contributions within the recent advancements in semantic caching and agentic workflows in database systems.

\subsection{Semantic Parsing and Natural Language Interfaces}

One major line of research in KGQA is semantic-parsing-based (SP-based) KGQA, which transforms natural language queries into logical forms to execute structured queries. SP-based approaches can be divided into multi-step and seq2seq methods. Multi-step approaches formulate this as a multi-step search problem, involving core entity identification and query graph expansion based on entity attributes and query predicates~\cite{yih2015semantic, yih2016value, luo2018knowledge, chen2019uhop, lanknowledge, lan2020query, oguz2022unik}. Seq2seq approaches treat semantic parsing as a translation task, leveraging fine-tuned language models to directly generate complete semantic expressions and retrieve relevant KG content~\cite{das2021case, ye2022rng, cao2022program, gu2022arcaneqa, shu2022tiara, hu2022logical, xie2022unifiedskg, yu2022decaf, zhang2023fc, gu2023don, xu2023fine}. For example, \cite{ye2022rng} generates and ranks candidate logical expressions, \cite{cao2022program} first creates a high-level sketch and then refines its arguments, and Decaf~\cite{yu2022decaf} introduces a retriever-reader-combiner structure that jointly generates answers and logical forms. While highly precise, these methods depend on strict schemas and exhibit low tolerance for parsing errors, struggling to handle the semantic ambiguities inherently present in complex QA.

\subsection{KG-based RAG and LLM-driven Execution}
\label{subsec:related-kgrag}
% LLM-based approach

Recently, the database and NLP communities have focused on LLM-based KGQA approaches that integrate semantic parsing and query generation into the reasoning process of LLMs~\cite{sunthink, ma2025thinkongraph, jiang2023structgpt}. StructGPT~\cite{jiang2023structgpt} designs a data retrieval interface, using prompts to enable LLMs to generate API or SPARQL calls. However, its direct tool-calling can blindly propagate planning errors. Beyond tool calling, ToG~\cite{sunthink} introduces beam search on KGs for iterative path exploration. ToG-2~\cite{ma2025thinkongraph} extends modality to joint KG-text reasoning with external Wikipedia.
Concurrent to our work, recent database research has started treating LLMs as core execution engines or query optimizers. For instance, SERAG\cite{liu2025serag} utilizes a continuously updated RAG vector database to help learned query optimizers avoid cold-start problems and learn from historical execution feedback. In contrast to these approaches, CacheRAG focuses on enhancing the LLM planner's continual learning for graph traversals via QA history based on relevance and diversity, while introducing strictly bounded depth/breadth expansions to significantly boost recall without sacrificing system stability. Two Meta KDD Cup 2024 solutions also developed LLM-based RAG: db3~\cite{xia2024winning} uses an LLM planner for routing, while apex~\cite{ouyang2024revisiting} crafts API extraction rules to match questions.

\subsection{Semantic Caching and Agentic Workflows}
\label{subsec_agentic_related}

Caching mechanisms have become indispensable for reducing latency and costs in modern LLM deployments. Recent data management literature, such as Cache-Craft~\cite{agarwal2025cache} and RAGCache~\cite{jin2025ragcache}, focuses on systems that optimize RAG workflows by efficiently caching precomputed Key-Value (KV) states or intermediate retrieved text chunks across similar queries. Unlike these infrastructure-level caches, CacheRAG proposes a higher-level semantic plan cache that stores logical reasoning paths and historical execution schemas to directly guide the LLM's multi-hop planning.

Regarding agentic workflows, systems like Buffer-of-Thoughts (BoT)~\cite{yang2024buffer} build a meta-buffer storing universal thought-templates distilled from problem-solving processes, while AFlow~\cite{zhangaflow} uses Monte Carlo Tree Search to generate general task workflows. These methods focus on free-form tasks, whereas CacheRAG targets KGQA governed by strict database schema constraints. Despite both using experience replay, CacheRAG differs fundamentally: 1) CacheRAG relies on unsupervised template generation (Auto Generation), offering higher generalizability to new KGs compared to supervised methods. 2) CacheRAG formulates cache retrieval as a submodular optimization problem, employing a two-layer index and MMR to balance relevance and structural diversity, whereas existing systems rely on naive semantic similarity. 3) CacheRAG's cached examples provide both reasoning logic and reusable physical KG triples, ensuring deterministic retrieval coverage.

In summary, CacheRAG is a highly generalizable cache-augmented KGQA system that enables continual learning, featuring unsupervised template generation, a submodular retrieval index, experience replay, and bounded two-dimensional KG path exploration.

\paragraph{Positioning: CacheRAG's Key Contributions.} 
CacheRAG introduces four novel system-level contributions that, while building on existing primitives (MMR, dense retrieval, graph traversal), constitute \textbf{the first principled integration of caching mechanisms into LLM-based KGQA}:

\textbf{(1) Cache-Augmented Continual Learning for KGQA.} 
We are the \textbf{first to systematically apply semantic caching to enable continual learning in LLM-based query planning}. Unlike traditional DB plan caches that optimize for query acceleration (LRU/LFU), our cache is designed to \textit{teach} the LLM recurring query patterns. Our ablation (Section~\ref{subsec_ablation}) presents up to 6\% performance gains, validating that \textit{learning from history fundamentally outperforms stateless planning}.

\textbf{(2) Provably Bounded Expansion Operators.} 
We introduce the \textbf{first formalization of bounded depth/breadth expansion} with strict complexity guarantees ($K_{depth} \le 3$, space complexity $O(K_{depth} \cdot K_{degree})$) for LLM-driven graph traversal. Unlike prior work that relies on unbounded LLM-guided exploration (risking OOM crashes), our design ensures \textit{production-ready deployment} while significantly improving retrieval recall.

\textbf{(3) Schema-Agnostic User Interface via ISR.} 
We propose a two-stage semantic parsing framework that \textbf{decouples user-facing natural language from physical KG schemas}, achieving 98.9\% attribute alignment precision while eliminating the need for users to learn SPARQL or predicate names.

\textbf{(4) Diversity-Aware Cache Retrieval via MMR.} 
We introduce MMR-based cache retrieval to maximize information gain in the LLM's limited context window. By balancing semantic relevance with structural diversity, our approach ensures each cached example contributes unique query pattern knowledge, preventing redundancy and mode collapse. 

%\paragraph{Positioning: What Makes CacheRAG Novel?} While individual components (MMR, dense retrieval, graph expansion) have precedents in isolation, \textbf{CacheRAG is the first work to systematically integrate them into a unified cache-augmented architecture for LLM-based KGQA}. Our novelty lies in three aspects: (1) \textbf{Problem formulation}: We are the first to identify and formalize the "stateless planner" problem in LLM-based KGQA, drawing an explicit analogy to database plan caching. (2) \textbf{Design principles}: We introduce LLM-specific adaptations (diversity-over-frequency, schema alignment, bounded execution) that differ fundamentally from traditional DB caching. (3) \textbf{Empirical validation}: Our component isolation ablation (Section~\ref{subsec_ablation}) provides validation that cache-guided planning outperforms blind expansion, answering a critical open question about where LLM-based retrieval gains originate.

%% file: secs/sec_problem.tex
\section{Problem Formalization}
\label{sec:problem}

\begin{figure*}[htbp]
  \centering
\includegraphics[width=0.8\linewidth]{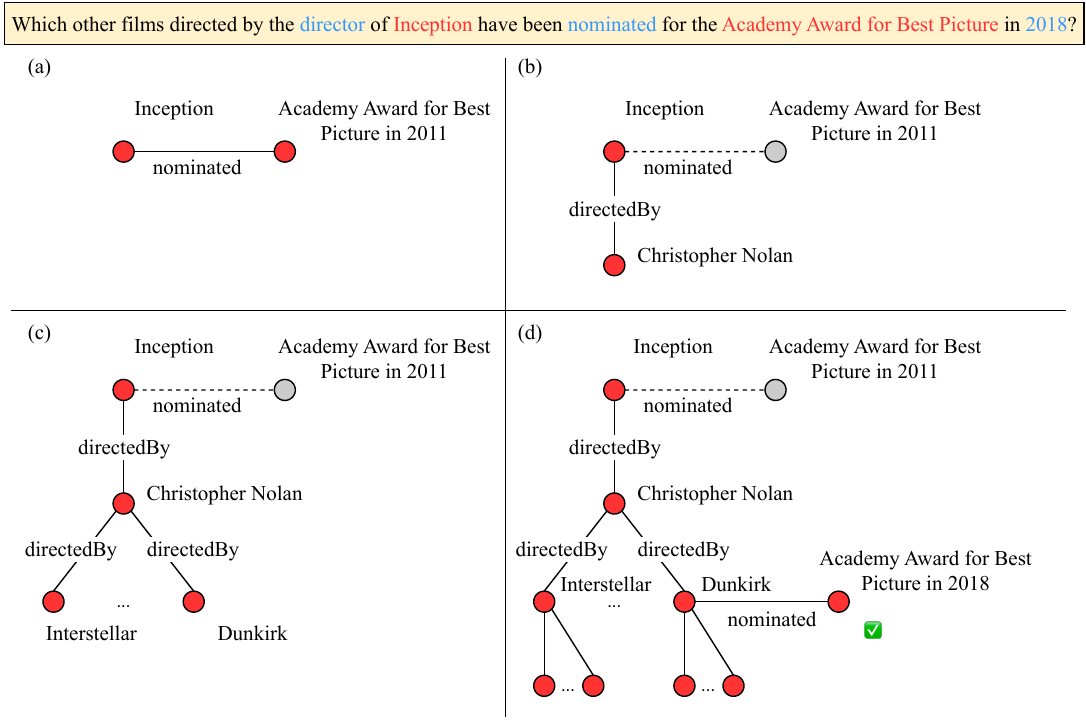}
\vspace{-1em}
      \caption{Illustration of retrieval path expansion for the running example. (a) Direct prompting: LLM incorrectly checks \texttt{nominated} attribute of ``Inception''. (b) With cache examples: LLM learns to identify the director via \texttt{directedBy} relation. (c) Bounded Depth Expansion: The $\sigma_{depth}$ operator safely chains to Christopher Nolan's other films. (d) Bounded Breadth Expansion: A Star-Pattern Scan ($\sigma_{breadth}$) retrieves 1-hop neighbors of ``Dunkirk'', successfully locating the 2018 Academy Award nomination.}
  \label{fig:expansion}
  \vspace{-1em}
\end{figure*}

\subsection{Core Definitions}

\noindent\textbf{Definition 1 (Knowledge Graph).} A Knowledge Graph is formally defined as a tuple $\mathcal{K} = (E, R, T)$, where:
\begin{itemize}
\item $E$: Finite set of entities (e.g., \textit{Inception}, \textit{Christopher Nolan})
\item $R \subseteq E \times P \times E$: Set of triples $(t_1, p, t_2)$, where $t_1, t_2 \in E$, and $p \in P$ is a predicate from the schema $\mathcal{S}$
\item $T$: Optional temporal annotations (for time-sensitive KGs like Wikidata)
\end{itemize}

\noindent\textbf{Definition 2 (Intermediate Semantic Representation).} To decouple user inputs from physical schema, we define an ISR as a tuple:
\[
\mathcal{I} = (e_{\text{raw}}, \Phi_{\text{raw}}, d). 
\]
Here, $e_{\text{raw}}$ is the raw topic entity, $\Phi_{\text{raw}}$ is a set of semantic constraints extracted from the natural language question $Q_{\text{NL}}$, and $d$ is the domain hint.

\noindent\textbf{Definition 3 (Query Plan).} A query plan is a sequence of retrieval operations:
\[
\pi = \langle op_1, op_2, ..., op_m \rangle, \quad op_i \in \{\text{SPARQL}, \text{API call}\}
\]
Executing $\pi$ on $\mathcal{K}$ yields a subgraph $G_\pi \subseteq \mathcal{K}$.

\noindent\textbf{Definition 4 (Semantic Cache).} The cache $\mathcal{C}$ stores successful retrieval histories:
\[
\mathcal{C} = \{(Q_i, \pi_i, A_i)\}_{i=1}^{N}, \quad Q_i \in \text{NL}, \pi_i \text{: query plan}, A_i \text{: verified answer}
\]
Each entry is indexed by domain $d$ and aspect attribute $a$: $\mathcal{C}[d][a]$.

\subsection{Problem Statement}

\noindent\textbf{Input:}
\begin{itemize}
\item A KG $\mathcal{K}$ with schema $\mathcal{S} = (P, \mathcal{O})$ (predicates + ontology)
\item A natural language question $Q_{\text{NL}}$
\item A semantic cache $\mathcal{C}$ (initially empty or auto-generated)
\end{itemize}

\noindent\textbf{Goal:} Design a system that:
\begin{enumerate}
\item Extracts a schema-agnostic ISR $\mathcal{I}$ from $Q_{\text{NL}}$
\item Retrieves diverse historical plans $S^* \subseteq \mathcal{C}$ (via MMR) to guide the LLM planner
\item Generates an executable query plan $\pi$ via in-context learning with $S^*$
\item Dynamically adjusts $\pi$ (depth/breadth expansion) if initial retrieval $G_\pi$ is insufficient
\item Returns a verified answer $A$ and updates $\mathcal{C} \gets \mathcal{C} \cup \{(Q_{\text{NL}}, \pi, A)\}$
\end{enumerate}

\noindent\textbf{Optimization Objectives:}
\begin{itemize}
\item \textbf{Accuracy:} Maximize $\mathbb{P}(A = A_{\text{gold}})$ (answer correctness)
\item \textbf{Efficiency:} Minimize $|\pi|$ (number of queries) while maintaining high recall
\item \textbf{Diversity:} Maximize information gain $\mathcal{H}(S^*)$ in retrieved cache examples
\end{itemize}

\noindent\textbf{Key Constraints:}
\begin{itemize}
\item \textbf{Schema Fidelity:} All predicates in $\pi$ must satisfy $p \in \mathcal{S}$ (no hallucination)
\item \textbf{Bounded Execution:} Depth expansion limited to $K_{\text{depth}} \leq 3$ hops, breadth expansion to Top-$K_{\text{degree}}$ neighbors
%\item \textbf{Cache Capacity:} $|\mathcal{C}| \leq B$ (bounded memory), eviction policy
\end{itemize}

\paragraph{System Design Rationale.} 
To realize the aforementioned optimization objectives, the system must organically resolve the four inherent execution bottlenecks introduced in Section~\ref{sec:intro} (i.e., schema hallucination, cache redundancy, incomplete retrieval, and cold-start degradation). We achieve this through the CacheRAG pipeline, an end-to-end architecture detailed in the following section.

%\subsection{Key Challenges}

%\paragraph{Challenge 1: Schema Hallucination.} The LLM may generate predicates $p \notin \mathcal{S}$, causing execution failures. \textit{Solution:} Backend adapter with local schema grounding (Section~\ref{subsec:semantic_parsing}).

%\paragraph{Challenge 2: Cache Redundancy.} Naive similarity-based retrieval yields low-diversity $S$, wasting LLM context. \textit{Solution:} MMR-based retrieval with hierarchical index (Section~\ref{subsec:diversity_cache}).

%\paragraph{Challenge 3: Incomplete Retrieval.} Single-hop plans $\pi$ may miss multi-hop answers. \textit{Solution:} Bounded expansion (Section~\ref{subsec:expansion}).

%\paragraph{Challenge 4: Cold Start.} Empty cache $\mathcal{C} = \emptyset$ degrades initial performance. \textit{Solution:} Auto-generation (Section~\ref{subsec:auto_gen}).

%% file: secs/sec_method.tex
\section{Methodology}
\label{sec_method}

\begin{figure}[t!]
  \centering
\includegraphics[width=0.9\columnwidth]{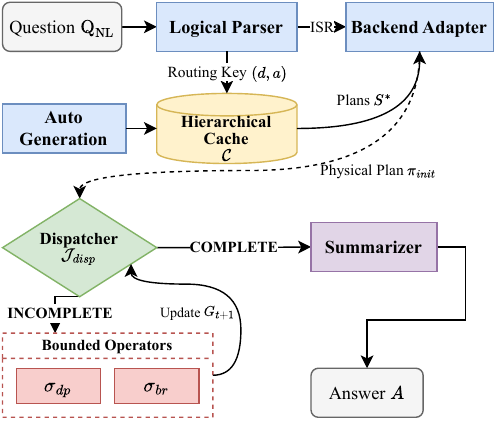}
\vspace{-1em}
      \caption{The overall pipeline of CacheRAG, featuring a dual-layer architecture: logical parsing/compilation (top) and heuristic execution with bounded operators (bottom).}
      \vspace{-1em}
  \label{fig:pipeline}
\end{figure}

The CacheRAG pipeline systematically transforms LLM-based reasoning into a rigorous, bounded database execution workflow. Rather than treating execution bottlenecks in isolation, our key novelty lies in the synergistic integration of four components: (1) schema-agnostic semantic parsing via an adapter pattern (Section~\ref{subsec:semantic_parsing}), (2) diversity-optimized hierarchical caching (Section~\ref{subsec:diversity_cache}), (3) deterministically bounded expansion operators (Section~\ref{subsec:expansion}), and (4) an offline auto-generation module for cold-start mitigation (Section~\ref{subsec:auto_gen}).
To clarify the design choices and structural logic, we illustrate our core modules while grounding them in our running example (from Figure~\ref{fig:expansion}): ``Which other films directed by the director of Inception have been nominated for the Academy Award for Best Picture in 2018?''

%To clarify the design choices and structural logic, we illustrate our core modules while grounding them in our running example (from Figure~\ref{fig:expansion}): ``Which other films directed by the director of Inception have been nominated for the Academy Award for Best Picture in 2018?''

\paragraph{System Architecture Overview.}
Figure~\ref{fig:pipeline} illustrates the overall architecture of CacheRAG, which operates as a continuous, stateful execution pipeline (formalized in Algorithm~\ref{algo:overall}). As depicted, the workflow is strictly divided into a logical planning phase and a physical execution phase. Given a natural language question $Q_{\text{NL}}$, the \textbf{Logical Parser} first extracts an Intermediate Semantic Representation (ISR). This ISR serves dual purposes: it acts as a routing key for the \textbf{Hierarchical Cache} to fetch structurally diverse historical plans, and provides the logical constraints for the \textbf{Backend Adapter} to compile an initial physical plan $\pi_{\text{init}}$. During the execution phase, the \textbf{Heuristic Dispatcher} evaluates the retrieved subgraph. If incomplete, it triggers the \textbf{Bounded Subgraph Operators} ($\sigma_{depth}$ and $\sigma_{breadth}$), iteratively updating the subgraph via a feedback loop until a termination state is reached, at which point the \textbf{Summarizer} generates the final answer $A$. We detail these modules in the following subsections.

\begin{algorithm}[t!]
\caption{Overall CacheRAG Architecture Process}
\label{algo:overall}
\begin{flushleft}
\textbf{Function Name:} CacheRAG\_Pipeline\\
\textbf{Input:} User question $Q_{\text{NL}}$, query time $t$\\
\textbf{Output:} Natural language answer $A$\\
1. Initialize the semantic cache $\mathcal{C}$ via Auto-Generation (Sec~\ref{subsec:auto_gen}).\\
2. $\mathcal{I} \gets \text{LogicalParser}(Q_{\text{NL}})$ \textit{\# Extract domain $d$, constraints $\Phi_{\text{raw}}$ (Sec~\ref{subsec:semantic_parsing})}\\
3. $S^* \gets \text{CacheRetrieval}(Q_{\text{NL}}, \mathcal{C}, \lambda, k, d, a)$ \textit{\# Fetch diverse plans (Sec~\ref{subsec:diversity_cache})}\\
4. $\pi_{\text{init}} \gets \text{BackendAdapter}(\mathcal{I}, \mathcal{S}_{local}, S^*)$ \textit{\# Compile physical plan}\\
5. $G_{\text{init}} \gets \text{Execute}(\pi_{\text{init}})$ \textit{\# Retrieve initial subgraph}\\
6. \textbf{if} $\mathcal{J}_{\text{dispatcher}}(Q_{\text{NL}}, t, G_{\text{init}}, \mathcal{C}) == \text{COMPLETE}$ \textbf{then}:\\
\quad 6.1 $A \gets \text{Summarize}(Q_{\text{NL}}, t, G_{\text{init}})$\\
\quad 6.2 $\text{CacheUpdate}(\mathcal{C}, Q_{\text{NL}}, \pi_{\text{init}}, A)$ \textit{\# Trigger eviction if $|\mathcal{C}| > B$}\\
\quad 6.3 \textbf{return} $A$\\
7. \textbf{else}: \textit{\# Insufficient retrieval, trigger heuristic expansion}\\
\quad 7.1 $\pi_{\text{final}}, G_{\text{final}} \gets \text{HeuristicTraversal}(Q_{\text{NL}}, t, \pi_{\text{init}}, G_{\text{init}}, \mathcal{C})$ \textit{\# (Sec~\ref{subsec:expansion})}\\
\quad 7.2 \textbf{if} $\mathcal{J}_{\text{dispatcher}}(Q_{\text{NL}}, t, G_{\text{final}}, \mathcal{C}) == \text{COMPLETE}$ \textbf{then}: \\
\quad \quad 7.2.1 $A \gets \text{Summarize}(Q_{\text{NL}}, t, G_{\text{final}})$\\
\quad \quad 7.2.2 $\text{CacheUpdate}(\mathcal{C}, Q_{\text{NL}}, \pi_{\text{final}}, A)$\\
\quad 7.3 \textbf{else}: $A \gets \text{LLM\_Fallback}(Q_{\text{NL}}, t)$ \textit{\# Answer via LLM directly}\\
\quad 7.4 \textbf{return} $A$
\end{flushleft}
\end{algorithm}

\subsection{Schema-Constrained Semantic Parsing via Adapter Pattern}
\label{subsec:semantic_parsing}

%\paragraph{Two-Stage Compilation Architecture.}
\paragraph{Our Design: Two-Stage Compilation Architecture.}
To address the severe bottlenecks of schema hallucination and lack of robustness inherent in zero-shot LLM prompting (as outlined in Section~\ref{sec:intro}), CacheRAG conceptualizes semantic parsing as a two-stage process. As depicted in the top logical layer of Figure~\ref{fig:pipeline}, this strictly decoupled design is analogous to the compilation of a logical query plan into a physical query plan in traditional DBMS.
%To address this, CacheRAG conceptualizes semantic parsing as a two-stage process. As depicted in the top logical layer of Figure~\ref{fig:pipeline}, this is analogous to the compilation of a logical query plan into a physical query plan in traditional DBMS.

\noindent\textbf{Stage 1: Schema-Agnostic ISR Extraction (Logical Plan).} The \textbf{Logical Parser} first acts purely as a syntactic parser, extracting an Intermediate Semantic Representation (ISR) from the user's natural language query. The ISR consists of:
\begin{itemize}
\item Topic entity $e_{\text{raw}}$ (e.g., ``Inception'')
\item Semantic constraints $\Phi_{\text{raw}} = \{\phi_1, \phi_2, ...\}$ (e.g., ``directed by'', ``Academy Award'', ``2018'')
\item Domain hint $d$ (e.g., ``movies''). To ensure high precision in multi-domain environments, we augment this extraction with a dynamic KG description cache (detailed in Appendix~\ref{subsec_multi}), enabling the parser to iteratively refine its domain routing accuracy based on execution feedback.
\end{itemize}

Critically, the ISR extraction prompt \textit{explicitly prohibits} the LLM from generating KG-specific predicate names, ensuring the extraction remains independent of the physical schema:

\begin{block}{ISR Extraction Prompt (Schema-Agnostic)}
Extract the main entity and key constraints from the question. Do NOT generate KG predicates or schema-specific terms: use only the natural language expressions from the user's question.
Question: ``Which other films directed by the director of Inception have been nominated for the Academy Award for Best Picture in 2018?''
Your output format:
- Main entity: [entity name]
- Constraints: [list of semantic constraints in natural language]
\end{block}

\noindent\textbf{Stage 2: Schema-Aware Query Compilation (Physical Plan).} To bridge the gap between the platform-agnostic ISR and the underlying heterogeneous data sources, CacheRAG employs a deterministic \textbf{Backend Adapter}. Instead of forcing the LLM to blindly output SPARQL syntax, the adapter first fetches the \textit{local physical schema} (i.e., the valid 1-hop attributes and relations $\mathcal{S}_{local}$ associated with the retrieved topic entity $e_{\text{raw}}$). 

The LLM is then repurposed as a \textbf{Query Compiler}. It receives both the logical constraints $\Phi_{\text{raw}}$ and the valid local schema $\mathcal{S}_{local}$ as context, and is tasked with grounding the natural language constraints to the exact physical predicates to assemble the executable API call or SPARQL query. By strictly decoupling the natural language reasoning phase from the physical graph execution, and forcing the LLM to select only from provided valid predicates, this design effectively prevents schema hallucination.

\paragraph{Running Example.}
\begin{itemize}
\item \textbf{Logical ISR:} Constraint ``directed by''
\item \textbf{Adapter Context:} System fetches local schema for ``Inception'': $\mathcal{S}_{local} =$ \{\texttt{directedBy}, \texttt{starring}, \texttt{genre}, \texttt{releaseDate}\}
\item \textbf{Physical Compilation:} The Backend Adapter maps the logical ``directed by'' to the physical \texttt{directedBy} and formats the exact executable SPARQL queries or valid API calls.
\end{itemize}

\paragraph{Error Handling and Robustness.} On CRAG dataset~\cite{yang2024crag}, this two-stage decoupling achieves:
\begin{itemize}
\item \textbf{98\% entity extraction accuracy} on CRAG (errors stem primarily from pronoun ambiguity, e.g., ``his latest work'', without conversational context).
\item \textbf{98.9\% physical compilation executable rate}, confirming that providing local schema context significantly reduces API execution failures.
\end{itemize}
If the LLM fails to map a constraint to the local schema, the system deterministically triggers \textbf{Breadth Expansion} (Section~\ref{subsec:expansion}) to explore multi-hop schema paths, gracefully handling schema ambiguity.

\subsection{Diversity-Aware Plan Caching and Retrieval Optimization}
\label{subsec:diversity_cache}

To provide the LLM with the ability to perform continual learning without weight updates, CacheRAG maintains a cache $\mathcal{C}$ of historical query plans. However, a critical design challenge arises during retrieval: naive semantic similarity search (e.g., standard KNN) tends to retrieve highly homogeneous queries. For instance, if the cache returns five structurally identical historical plans asking about ``movie awards'', the marginal information gain for the LLM is near zero. This redundancy wastes the limited context window and induces reasoning homogeneity (mode collapse).

To address this, CacheRAG must enforce structural diversity during cache retrieval while maintaining strict online latency bounds. Evaluating a diverse subset from a global cache pool of size $N$ naively requires $\mathcal{O}(N^2)$ pairwise comparisons. We bypass this bottleneck through a two-stage retrieval architecture: a hierarchical index for search space pruning, followed by a Maximal Marginal Relevance (MMR) mechanism for diversity-aware selection.

\begin{figure}[hbtp]
  \centering
  \includegraphics[width=0.9\linewidth]{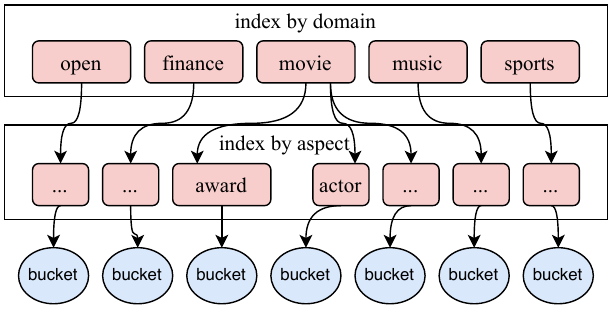}
  \vspace{-1em}
  \caption{The two-layer cache structure for multi-domain KGs (e.g., CRAG dataset). The first layer indexes by domain (movies, music, sports), while the second layer partitions by aspect attributes (award, cast, director). For single-domain KGs like Freebase (CWQ), we simplify to a one-layer index based on aspect only.}
  \vspace{-1em}
  \label{fig:cache-structure}
\end{figure}

\paragraph{Stage 1: Hierarchical Indexing for Hard Pruning.} 
To seamlessly bridge the logical parsing phase with cache retrieval, CacheRAG utilizes the extracted ISR $\mathcal{I}$ as the routing key (depicted as $(d, a)$ in Figure~\ref{fig:pipeline}). As further detailed in Figure~\ref{fig:cache-structure}, we design a two-layer routing structure to organize the cache:
\begin{itemize}[leftmargin=*]
\item \textbf{Layer 1 (Domain):} Routed by the domain hint $d$ from the ISR, this layer separates examples by semantic domains (e.g., movies, music, sports for the CRAG dataset). This strict boundary prevents cross-domain contamination, ensuring that examples from unrelated domains do not dilute the in-context learning relevance.
\item \textbf{Layer 2 (Aspect Attribute):} Further partitions examples by the core query intent, which is abstracted directly from the semantic constraints $\Phi_{\text{raw}}$. For instance, under the ``movie'' domain, queries with constraints regarding nominations (abstracted as the ``award'' aspect) are isolated from those inquiring about actor relations (the ``cast'' aspect). For single-domain KGs (e.g., Freebase in CWQ), this degrades gracefully to a one-layer aspect index.
\end{itemize}
This hierarchical structure acts as a hard search space pruning step. Guided deterministically by the logical ISR, it restricts the subsequent diversity evaluation from the global cache to a highly localized bucket, bounding the search complexity to $\mathcal{O}(b \log b)$, where $b = |\mathcal{C}_{bucket}| \ll N$ denotes the size of the target bucket.

\paragraph{Stage 2: MMR-based Soft Diversity Enforcement.}
Within the target bucket, CacheRAG implements the Maximal Marginal Relevance (MMR) strategy (detailed in Algorithm~\ref{algo:cache}) to select candidate plans. At each step, we iteratively select a candidate cached entry $c_j = (Q_j, \pi_j, A_j)$ that maximizes the marginal gain between its semantic relevance to the input query $Q_{\text{NL}}$ and its diversity against the already selected set $S^*$:
\begin{equation}
\small
\arg\max_{c_j \in \mathcal{C}_{bucket} \setminus S^*} \Big( \lambda \cdot \text{Sim}(Q_{\text{NL}}, Q_j) - (1 - \lambda) \cdot \max_{c_k \in S^*} \text{Sim}(Q_j, Q_k) \Big)
\end{equation}
By dynamically balancing relevance and diversity via the penalty parameter $\lambda$, this step deliberately avoids redundant token consumption and formulates a structurally heterogeneous $k$-shot prompt for the Backend Adapter.

\begin{algorithm}[t!]
\caption{Diversity-Aware Cache Retrieval via MMR}
\label{algo:cache}
\begin{flushleft}
\textbf{Function Name:} CacheRetrieval\\
\textbf{Input:} Input query $Q_{\text{NL}}$, semantic cache $\mathcal{C}$, penalty parameter $\lambda$, sample size $k$, domain $d$, aspect $a$ \\
\textbf{Output:} Selected historical plans $S^*$ for compiler context\\
1. $\mathcal{C}_{bucket} \gets \mathcal{C}[d][a]$ \textit{\# Hard pruning: Fetch from localized index}\\
2. \textbf{if} $|\mathcal{C}_{bucket}| < k$ \textbf{then} $\mathcal{C}_{bucket} \gets \mathcal{C}[d][*]$  \textit{\# Fallback: Relax to domain level}\\
3. $S^* \gets \emptyset$  \textit{\# Initialize selected subset}\\
4. \textbf{while} $|S^*| < k$ \textbf{and} $\mathcal{C}_{bucket} \neq \emptyset$ \textbf{do}: \\
\quad 4.1 $c^* \gets \underset{c_j \in \mathcal{C}_{bucket}}{\arg\max}\ \big[\lambda \cdot \text{Sim}(Q_{\text{NL}}, Q_j)$\\
\qquad\quad $- (1{-}\lambda) \cdot \max_{c_k \in S^*} \text{Sim}(Q_j, Q_k)\big]$ \\
\quad 4.2 $S^* \gets S^* \cup \{c^*\}$ \\
\quad 4.3 $\mathcal{C}_{bucket} \gets \mathcal{C}_{bucket} \setminus \{c^*\}$ \\
5. \textbf{return} $S^*$
\end{flushleft}
\end{algorithm}

\paragraph{Semantic Cache Storage and Bounded Memory Extension.}
By default, CacheRAG maintains an unbounded cache to accumulate all successful historical plans, maximizing the repository for continual learning. Because our two-layer hierarchical index proactively restricts the search space to highly localized buckets, the computational overhead of MMR retrieval remains strictly bounded ($\mathcal{O}(b \log b)$) even as the global cache size $N$ grows indefinitely. 

However, for practical deployments with strict hardware constraints, CacheRAG can be seamlessly extended with a bounded capacity limit $B$ per bucket. In such memory-constrained scenarios, applying a standard Least Recently Used (LRU) eviction policy is sufficient to maintain a high-quality repository. Traditional frequency-based eviction does not degrade the structural diversity here, as the hierarchical index has already rigidly partitioned the query intents prior to eviction. This LRU-based extension is empirically validated to save 64\% of storage space with a negligible ($<2\%$) performance drop (as detailed in Section~\ref{subsec_lru}), proving the robustness of our architecture under bounded memory footprints.

\paragraph{Justification \& Running Example.} 
Consider processing our running example (``Which other films directed by the director of Inception have been nominated for the Academy Award for Best Picture in 2018?''). The system:
\begin{enumerate}
\item Routes the query to the $\mathcal{C}[\text{movies}][\text{award}]$ bucket via the two-layer index.
\item Applies MMR (Algorithm~\ref{algo:cache}) to fetch structurally diverse semantic plans, such as:
\begin{itemize}
\item A multi-hop plan finding related entities via a shared person (director $\rightarrow$ other movies).
\item A plan demonstrating temporal filtering logic (year $= 2018$).
\item A complex specific query structure resolving ``Academy Award for Best Picture''.
\end{itemize}
\end{enumerate}
The MMR penalty systematically restricts the system from fetching redundant ``movie nomination'' plans. Instead, it compiles a diverse set of plans that comprehensively teach the LLM the exact compositional logic needed to assemble the final complex multi-hop execution path.

\subsection{Dynamic Plan Adjustment via Bounded Subgraph Operators}
\label{subsec:expansion}

When the initial physical query plan $\pi_{\text{init}}$ is executed, the retrieved subgraph might be insufficient to answer complex queries. Rather than allowing unconstrained LLM reasoning, which frequently leads to combinatorial explosion, unbounded API calls, or Out-Of-Memory (OOM) errors during deep graph traversal, CacheRAG formulates the dynamic plan adjustment as a set of \textbf{Bounded Heuristic Graph Operators}.

As shown in the lower execution layer of Figure~\ref{fig:pipeline}, the LLM is decoupled from direct graph navigation and instead acts strictly as a \textbf{Lightweight Heuristic Dispatcher} $\mathcal{J}_{\text{dispatcher}}(Q_{\text{NL}}, t, G_t, \mathcal{C})$. Given the currently retrieved subgraph $G_t$, the dispatcher evaluates the semantic completeness of the current state. It then greedily routes the execution to one of two deterministic physical expansion operators or to an early termination state ($\mathcal{J} = \text{COMPLETE}$). To guarantee predictable latency and prevent context window overflow, system-level safeguards are enforced as hard execution boundaries during operator invocation.

\paragraph{Bounded Depth Expansion ($\sigma_{depth}$).} Conceptualized as a \textbf{Graph-based Index Nested Loop Join}, this operator extends the multi-hop path by exploring contexts along specific relational edges. The heuristic dispatcher selects the most promising relational hop based on the current frontier nodes. To prevent infinite loops and tightly bound the execution cost, the depth of this nested join is strictly constrained by a system parameter ($K_{depth} \le 3$). As illustrated in Figure~\ref{fig:expansion}(c), depth expansion enables the system to safely chain from ``Inception'' $\xrightarrow{\text{directedBy}}$ ``Christopher Nolan'' $\xrightarrow{\text{directed}}$ [``Interstellar'', ``Dunkirk'', ...].

\paragraph{Bounded Breadth Expansion ($\sigma_{breadth}$).} When targeted reasoning fails due to implicit schema mismatches, CacheRAG executes a \textbf{Star-Pattern Neighborhood Scan} to broadly explore the 1-hop neighborhood $\mathcal{N}(v)$ of bottleneck entities. To prevent memory explosion on super-nodes, we apply a Top-K pruning strategy (e.g., via dense similarity), ensuring the neighborhood size never exceeds an administrative bound $K_{degree}$. Consequently, the space complexity of the traversal is strictly maintained at $\mathcal{O}(K_{depth} \cdot K_{degree})$. Figure~\ref{fig:expansion}(d) shows how breadth expansion retrieves all neighbors of ``Dunkirk'', capturing the 2018 Academy Award nomination that would be missed by targeted queries alone. 

Both operators iteratively feed the expanded results back into the dispatcher (depicted as the ``Update $G_{t+1}$'' feedback loop in Figure~\ref{fig:pipeline}) until the information is semantically complete.

\begin{algorithm}[t!]
\caption{Heuristic Dispatching for Bounded Subgraph Expansion}
\label{algo:expansion}
\begin{flushleft}
\textbf{Function Name:} HeuristicTraversal\\
\textbf{Input:} User question $Q_{\text{NL}}$, query time $t$, current physical plan $\pi$, retrieved subgraph $G_t$, semantic cache $\mathcal{C}$ \\
\textbf{Output:} Extended physical plan $\pi_{\text{final}}$, Expanded subgraph $G_{\text{final}}$\\
1. $\text{state\_status} \gets \mathcal{J}_{\text{dispatcher}}(Q_{\text{NL}}, t, G_t, \mathcal{C})$ \textit{\# Evaluate logical completeness}\\
2. \textbf{if} $\text{state\_status} == \text{INCOMPLETE}$ \textbf{and} $|\pi| < K_{depth}$ \textbf{then}: \\
\quad 2.1 $r_{next}, V_{frontier} \gets \text{ExtractJoinCondition}(\mathcal{J}_{\text{dispatcher}})$ \\
\quad 2.2 $G_{depth} \gets \sigma_{depth}(V_{frontier} \bowtie r_{next})$ \textit{\# Execute Index Nested Loop Join}\\
\quad 2.3 $G_{breadth} \gets \sigma_{breadth}(\mathcal{N}(V_{frontier}), K_{degree})$ \textit{\# Execute Bounded Star Scan}\\
\quad 2.4 $G_{t+1} \gets G_t \cup G_{depth} \cup G_{breadth}$ \textit{\# Update sub-graph state}\\
\quad 2.5 $\pi_{\text{next}} \gets \pi \cup \{r_{next}\}$ \textit{\# Append execution path}\\
\quad 2.6 \textbf{return} HeuristicTraversal($Q_{\text{NL}}, t, \pi_{\text{next}}, G_{t+1}, \mathcal{C}$) \textit{\# Recursive routing}\\
3. \textbf{else}: \\
\quad \textit{\# Reached COMPLETE state or hit hard execution bound $K_{depth}$}\\
\quad 3.1 \textbf{return} $\pi, G_t$
\end{flushleft}
\end{algorithm}

\paragraph{Justification \& Running Example.} Figure~\ref{fig:expansion} illustrates the complete heuristic expansion process for our running example:

\begin{enumerate}
\item \textbf{Initial Execution (Figure~\ref{fig:expansion}(a)):} Without cache guidance, the system generates a logical plan to query the \texttt{nominated} attribute directly for ``Inception'' in 2018. Since Inception was nominated in 2011, this returns an empty subgraph. The dispatcher evaluates the state as incomplete.

\item \textbf{Cache-Guided Refinement (Figure~\ref{fig:expansion}(b)):} With cached semantic plans from $\mathcal{C}[\text{movies}][\text{award}]$, the system compiles the correct initial topology: first identify the director via \texttt{directedBy}. However, this single-hop path still lacks the final answer.

\item \textbf{Depth Expansion (Figure~\ref{fig:expansion}(c)):} Algorithm~\ref{algo:expansion} routes execution to the $\sigma_{depth}$ operator, chaining from ``Christopher Nolan'' to his filmography: [``Interstellar'', ``Dunkirk'', ``Tenet'', ...]. 

\item \textbf{Breadth Expansion (Figure~\ref{fig:expansion}(d)):} Simultaneously, the $\sigma_{breadth}$ operator executes bounded star scans for the retrieved entities. For ``Dunkirk'', this safely captures its local topology, including the ``Academy Award for Best Picture'' (2018) node. The dispatcher evaluates the subgraph as complete and terminates the traversal.
\end{enumerate}

This architecture guarantees maximum retrieval coverage for complex multi-hop QA. The strict deterministic bounds ($K_{depth} \le 3$, $K_{degree}$ pruning) proactively prevent catastrophic execution scenarios (e.g., unbounded API calls or OOM errors) common in unconstrained LLM agents. Meanwhile, the cache-guided heuristic dispatching significantly improves reasoning accuracy compared to static, single-round planning (as demonstrated in Section~\ref{subsec_ablation}).

\subsection{Cache Pre-Warming via Auto-Generation}
\label{subsec:auto_gen}

As discussed in Section~\ref{sec:problem}, deploying CacheRAG with an entirely empty memory degrades the system to a stateless planner during the initial queries. To alleviate this cold-start bottleneck, we implement an offline Auto-Generation module to populate the semantic cache $\mathcal{C}$ with fundamental operational heuristics prior to deployment. Rather than relying on human annotations, this process acts as an automated exploratory view materialization, executed in three straightforward steps:

\paragraph{1. Star-Schema Sampling and Synthesis.}
We uniformly sample entities along with their 1-hop relational neighbors (i.e., local star-schemas consisting of valid KG triples). Instead of assuming these shallow structures can only answer simple queries, we leverage an LLM to synthesize diverse natural language questions $Q_{\text{NL}}$ based on these subgraphs. Crucially, as shown in our prompt skeleton, the LLM is explicitly instructed to generate not only single-fact questions but also compositional queries requiring multi-triple reasoning (e.g., comparison and summarization). 

\begin{block}{Auto-Generation Prompt Skeleton}
You are an expert in question generation. Your task is to generate reasonable natural language questions based on the provided KG triples.
- Generate up to 5 questions that humans might naturally ask. 
- Avoid meaningless ID-based or purely numerical questions (Respond NA if unavoidable).
- **Crucial:** Generate both simple questions (1 triple) AND complex questions (e.g., comparison and summarization involving multiple triples).
Output format: [QUESTION], [TRIPLES], [ANSWER]
Current triples: [SAMPLED TRIPLES]
\end{block}

\paragraph{2. Automated Filtering and Compilation.}
To ensure the high quality of the materialization, the LLM acts as a secondary evaluator to filter out illogical or unanswerable generated pairs. For the surviving valid tuples, the system deterministically compiles the selected `[TRIPLES]' into executable API calls or SPARQL queries, forming the physical plan $\pi$. The resulting $\langle Q_{\text{NL}}, \pi, A \rangle$ tuples are ingested into the hierarchical cache to serve as the initial in-context learning baseline.

\paragraph{3. Handling Zero-Similarity and Online Growth.}
A natural limitation of offline random sampling is that it cannot guarantee complete structural coverage of all possible user intents. If a novel online query exhibits near-zero similarity with all pre-warmed examples (i.e., a severe cache miss), the system handles this cold-start gracefully. Under the MMR formulation, if the relevance score falls below a minimum threshold, the Query Compiler receives an empty context $S^* = \emptyset$ and degrades to zero-shot execution relying entirely on the local schema $\mathcal{S}_{local}$ (Section~\ref{subsec:semantic_parsing}). 

To prevent execution failure in this state, the system deterministically falls back on our Bounded Subgraph Operators ($\sigma_{depth}$ and $\sigma_{breadth}$ detailed in Section~\ref{subsec:expansion}) to broadly explore the graph. Once this heuristic traversal successfully locates the answer, the newly discovered execution path is dynamically added to $\mathcal{C}$. Thus, the auto-generation module only needs to provide a foundational ``warm start''; the cache subsequently grows and adapts organically to long-tail queries through regular online interactions.

%% file: secs/sec_experiment.tex
\section{Experiments}
\label{sec_experiments}
We present the experimental datasets (Section~\ref{subsec_datasets}), metrics (Section~\ref{subsec_metric}), baselines (Section~\ref{subsec_baseline}), and configurations (Section~\ref{subsec_config}). Our evaluation includes the following aspects: (1) comparison against the baselines on the CRAG dataset (Section~\ref{subsec_results}), (2) generalizability of CacheRAG on SPARQL-based datasets (Section~\ref{subsec_general}), (3) ablation study (Section~\ref{subsec_ablation}), (4) robustness of LLM backbones (Section~\ref{subsec_robust}), (5) robustness of two-layer indexing (Section~\ref{subsec_consistency}), (6) efficiency (Section~\ref{subsec_efficiency}), (7) scalability (Section~\ref{subsec_scalability}), (8) parameter sensitivity (Section~\ref{subsec_para}), (9) detailed head, torso, tail analysis (Section~\ref{subsec_htt}), and (10) analysis on KG domain routing (Section~\ref{subsec_multi}).

\subsection{Experimental Datasets}
\label{subsec_datasets}
We consider the widely used CRAG~\cite{yang2024crag}, QALD-10-en~\cite{Usbeck2023QALD10T}, WebQSP~\cite{yih2016value}, and CWQ~\cite{talmor-berant-2018-web} datasets in our experiments. Details are presented in Section~\ref{subsec_datasets_details}.

\subsection{Metrics and Evaluation Scheme}
\label{subsec_metric}

We adopted the auto-evaluation scheme used in CRAG~\cite{yang2024crag} to evaluate the correctness of the answers on the CRAG dataset. According to~\cite{yang2024crag}, the detailed criterion for accurate, missing, and hallucination is defined as follows:
\begin{itemize}
    \item Correct: The response correctly answers the user's question and contains no hallucinated content, or the response provides a useful answer to the user's question but may contain minor errors that do not harm the usefulness of the answer
    \item Missing: The response is ``I don’t know'', ``I'm sorry I can't find ...'', a system error such as an empty response, or a request from the system to clarify the original question.
    \item Incorrect: The response provides wrong or irrelevant information to answer the user's question.
\end{itemize}

Following the settings of CRAG~\cite{yang2024crag}, we adopted the Llama-3.1-70B-Instruct model as the auto-evaluation critic to classify the answers into correct, missing, and incorrect based on the question and the ground truth answer. {The auto-evaluation critic achieves an overall reliability of 98.4\%, demonstrating the reliability of the auto-evaluation scheme (Details in Section~\ref{subsec_metric_details}).} We report the accuracy ($A$, correct answer rate), miss rate ($M$, missing answer rate), hallucination rate ($H$, incorrect answer rate), and additionally the {\textit truthfulness} score ($T$), {which is defined as $T=1*A+0*M-1*H$ (the difference between accuracy and hallucination rate), penalizing the cases where the methods hallucinate.}

As for the SPARQL-based datasets QALD-10-en, WebQSP, and CWQ, we followed their respective evaluation metric: the Hit@1 score. Full details of metrics and evaluation scheme are presented in Section~\ref{subsec_metric_details}

\subsection{Baselines}
\label{subsec_baseline}

We compare our approach against the following baselines, which can be categorized into LLM base models (\textit{GPT-4o~\cite{hurst2024gpt}}, \textit{Llama-3.1-70B-Instruct~\cite{dubey2024llama}}, and \textit{Deepseek-chat-V3-0324~\cite{liu2024deepseek}}), LLM tool-calling models (\textit{StructGPT~\cite{jiang2023structgpt}} variants), KDD Cup-winning solutions (\textit{db3~\cite{xia2024winning}} and \textit{apex~\cite{ouyang2024revisiting}}), and SPARQL-based state-of-the-art solutions (\textit{ToG~\cite{sunthink}}, \textit{ToG-2~\cite{ma2025thinkongraph}}, \textit{sparql-qa~\cite{borroto2022sparql}}, and \textit{Decaf~\cite{yu2022decaf}}). Full details are presented in Section~\ref{subsec_baseline_details}.

\subsection{Configurations}
\label{subsec_config}

All experiments were performed on four NVIDIA A100 GPUs (80GB). We used the Deepseek-chat-V3-0324 model as the base LLM for CacheRAG across the CRAG, QALD-10-en, WebQSP, and CWQ datasets. The default MMR penalty parameter $\lambda$ is set to 0.5. The sample size $k$ for the structurally diverse plans provided to the LLM Query Compiler is set to 5. The similarity measure ($\text{Sim}(\cdot)$) used in the MMR formulation is based on BM25. We apply Chain-of-Thought reasoning during the physical query compilation phase to facilitate complex logical mapping. 
For the multi-domain CRAG dataset, we additionally deploy a multi-KG domain router (based on the Llama-3.1-8B-Instruct model with temperature set to 0) to assist the Logical Parser in determining the domain hint $d$. For the single-domain SPARQL-based datasets (QALD-10-en, WebQSP, and CWQ), we utilize the \texttt{llama-7b-wikiwebquestions-qald7} model\footnote{https://huggingface.co/stanford-oval/llama-7b-wikiwebquestions-qald7}~\cite{xu2023fine} to aid the extraction of the raw topic entity $e_{\text{raw}}$ within the ISR. Since these datasets do not distinguish domains, we gracefully degrade the two-layer hierarchical index into a one-layer aspect index, storing all cached plans within the same global domain bucket. 
During the Bounded Breadth Expansion ($\sigma_{breadth}$), a Dense Passage Retrieval (DPR) model~\cite{karpukhin2020dense} is employed as the Top-$K$ pruning strategy to filter irrelevant schema predicates. Specifically, we compute the similarity between the natural language question $Q_{\text{NL}}$ and the 1-hop predicates, preserving the top-30 ($K_{degree} = 30$) predicates to strictly maintain bounded space complexity.
When adapting the system to a new KG, users only need to map specific information into the predefined slots of the Backend Adapter templates (e.g., injecting the user query into [$Q_{\text{NL}}$]). This schema-agnostic configuration is straightforward and requires no deep expertise in prompt engineering. We estimate that configuring these structural adapter templates typically requires only 10 to 20 minutes of manual effort per dataset, depending on the physical KG schema's complexity. These templates function as generalizable compilation blueprints, ensuring that CacheRAG's deployment is highly efficient and reproducible across heterogeneous data sources.

%{\color{red} TODO: details of config}

\subsection{Main Results}
\label{subsec_results}

We present the experimental results on the CRAG dataset in Table~\ref{tab:baseline_crag}. In general, CacheRAG achieves the best performance on three out of four evaluation metrics compared to the baselines. Specifically, CacheRAG outperforms the state-of-the-art models by 13.2\% and 17.5\% in terms of accuracy and truthfulness scores, alongside a dramatically lower miss rate (over 38\% lower than the StructGPT variants).

Regarding the modest increase in the hallucination metric (between 3\% and 6.5\%) compared to some baselines, it is crucial to distinguish between two types of hallucinations. As discussed in Section~\ref{subsec:semantic_parsing}, CacheRAG effectively eliminates \textit{schema hallucination} (i.e., generating non-existent graph predicates that cause execution failures), evidenced by our 98.9\% physical compilation executable rate. The slight increase observed in the evaluation metric reflects \textit{answer content hallucination} during the final LLM summarization phase. Because CacheRAG employs bounded depth and breadth expansions to retrieve significantly more comprehensive and long-tail multi-hop contexts, the final Summarizer faces a more complex synthesis task. Given the substantial 13.2\% gain in overall accuracy and the dramatic reduction in miss rate, this modest increase in summarization hallucination is a well-contained trade-off for resolving the fundamental bottleneck of incomplete retrieval.

We believe the reason why CacheRAG outperforms the state-of-the-art baselines is two-fold: 1) The implementation of a caching mechanism provides a strong reference for the LLM planner to learn from its past interactions. The LLM base models, LLM tool calling, and KDD Cup-winning solutions all ignore the importance of valuable QA experience. As a result, the planners in these methods can only generate retrieval plans based on the current context instead of continuously learning from relevant QA experiences. Our CacheRAG approach helps the LLM planner effectively learn from QA experiences and thus achieve higher accuracy and truthfulness scores compared to existing methods. 2) Our KG exploration method provides better retrieval coverage. Instead of enhancing the retrieval paths in both breadth and depth dimensions, existing baselines either ignore the KG content (LLM Base Models) or neglect exploration in breadth (LLM Tool Calling Models and KDD Cup Winning Solutions). Our solution implements additional breadth expansion in case the KG content retrieved after depth expansion is still not sufficient for QA, which provides better KG recall (the retrieval recall is improved from $0.756$ to $0.927$). As a result, our method presents significantly lower miss rates.

% struct(4o) + 5; struct(l) + 10

\begin{table}[hbtp]
  \caption{Main Experimental Results on CRAG dataset. CacheRAG improves over state-of-the-art by 13.2\% and 17.5\% in terms of accuracy and truthfulness score.}
  \vspace{-1em}
  \small
  \label{tab:baseline_crag}
\centering
\begin{tabular}{lrrrrr}
\toprule 
 \textbf{Model} & \textbf{Accu.} & \textbf{Hall.} & \textbf{Miss.} & \textbf{Truth.} \\
\midrule
 GPT-4o   &     0.341               &                 0.090       &   0.569                &   0.251             \\
 Llama   &      0.306              &   0.080                     & 0.614                  & 0.227               \\
  Deepseek & 0.400 & 0.140 & 0.460 & 0.259\\
 StructGPT (GPT-4o)~\cite{jiang2023structgpt}   &    0.415                &      \textbf{0.047}                  &    0.542               &     0.368           \\
 StructGPT (Llama)~\cite{jiang2023structgpt}   &      0.318              &           0.057             &     0.623              &       0.261         \\
 StructGPT (Deepseek)~\cite{jiang2023structgpt} & 0.475 & 0.080 & 0.445 & 0.395\\
 apex~\cite{ouyang2024revisiting}   &       0.692             &    0.156                    &    0.152               &       0.536         \\
 db3~\cite{xia2024winning}   &         0.555           &  0.176                      &             0.259      &         0.379       \\
\cmidrule{1-5}
 CacheRAG   &     \textbf{0.824}               &                0.112        & \textbf{0.064}                  & \textbf{0.711}                            \\
\bottomrule
\end{tabular}
\end{table}

\begin{table}[hbtp]
	\centering
	\caption{Experimental results on the QALD-10-en, WebQSP, and CWQ datasets. The prior fine-tuning SOTAs include the best-known fine-tuned methods on each dataset: $\alpha$: sparql-qa~\cite{borroto2022sparql}; $\beta$: Decaf~\cite{yu2022decaf}; $\gamma$: CBR~\cite{das2021case}. The baseline performance is directly taken from ToG and ToG-2. Unlike other methods, ToG-2 additionally links to external Wikipedia pages to obtain context information. CacheRAG outperforms all the state-of-the-art baselines on the three datasets.}
    \vspace{-1em}
    \small
	\label{tab:qald-results}
	\begin{tabular}{llll}
		\toprule
		\textbf{Model}             & \textbf{QALD-10} & \textbf{WebQSP}  & \textbf{CWQ} \\    
		\midrule
		Prior Finetuning SOTAs & 0.454$^{\alpha}$ & 0.821$^{\beta}$ & 0.704$^{\gamma}$\\
		ToG~\cite{sunthink} & 0.502 & 0.762  & 0.695\\
		ToG-2~\cite{ma2025thinkongraph} & 0.541 & 0.811  & - \\
		\midrule
		CacheRAG & \textbf{0.587} & \textbf{0.840} & \textbf{0.736}\\
		\bottomrule
	\end{tabular}
\end{table}

\subsection{Generalizability}
\label{subsec_general}

To comprehensively evaluate the generalizability of CacheRAG on other KGQA datasets, we adapted the CacheRAG method for SPARQL-based datasets: QALD-10-en, WebQSP, and CWQ datasets. Specifically, instead of prompting the CacheRAG model to generate API function calls, we prompted it to generate SPARQL queries and execute those queries to get relevant KG triples for summarization. We compared the CacheRAG method against the respective state-of-the-art methods on these datasets.

We present the experimental results in Table~\ref{tab:qald-results}. In general, CacheRAG outperforms all the state-of-the-art solutions on the three datasets. Specifically, CacheRAG improves the state-of-the-art performance by $4.6\%$, $1.9\%$, and $3.2\%$ on the QALD-10-en, WebQSP, and CWQ datasets, respectively. We are surprised to see that our approach outperforms ToG-2 and prior fine-tuning SOTAs, despite the fact that ToG-2 additionally relies on external Wikipedia pages and that our method does not include a fine-tuning step to improve summarization quality, which showcases the effectiveness of continual learning on QA records and the advantages of KG exploration in depth and breadth.

%{To investigate why our method outperforms the baseline, we further analyze the performance of several variants of CacheRAG on the SPARQL-based KGs and present the results in Table~\ref{tab:qald-analysis}. Specifically, we consider CacheRAG w/o exploration (remove depth and breadth exploration), CacheRAG w/o caching (remove cache), and CacheRAG prompt (adopt the same set of prompts used by ToG). We found that the KG exploration and caching components were most critical to maintaining CacheRAG's performance, with performance declining by an average of 5.6\% and 2.8\% respectively after their removal. In contrast, the impact of changing the prompt on the performance of CacheRAG is within 1\%.}

\begin{comment}
\begin{table}[hbtp]
	\centering
	\caption{{Detailed analysis.}}
    \small
    \vspace{-1em}
	\label{tab:qald-analysis}
	\begin{tabular}{llll}
		\toprule
		\textbf{Model}             & \textbf{QALD-10} & \textbf{WebQSP}  & \textbf{CWQ} \\    
		\midrule
		CacheRAG w/o exploration & 0.552 & 0.787 & 0.656\\
        CacheRAG w/o caching & 0.576 & 0.816 & 0.688 \\
        CacheRAG prompt & 0.584 & 0.832 & 0.728 \\
		\bottomrule
	\end{tabular}
    \vspace{-1em}
\end{table}
\end{comment}

\subsection{Ablation Study}
\label{subsec_ablation}

To fully evaluate the effect of each component of CacheRAG, we conducted an ablation study by comparing the performance of CacheRAG and the following variants.

\begin{itemize}
    \item {\textbf{CacheRAG w/o depth:} represents the variant that we do not perform depth expansion, we only perform single round retrieval based on the initial planning.}
    \item \textbf{CacheRAG w/o breadth:} represents the variant that we do not perform additional breadth expansion, we only perform depth expansion for KG retrieval paths.
    \item \textbf{CacheRAG w/o caching:} represents the variant that removes the dynamic semantic cache, forcing the Query Compiler to rely on a static set of few-shot examplar plans.
    \item {\textbf{CacheRAG w/o MMR:} represents the variant that we replace the MMR score as BM25.}
    \item {\textbf{CacheRAG one layer:} represents the variant that we replace the two-layer index as one-layee domain index.}
     \item \textbf{CacheRAG w/o auto-generation:} represents the variant that we consider empty example cache to start.
    %\item \textbf{CacheRAG w/o domain:} represents the variant that we do not consider domain routing, instead, we directly provide all API function calls on all domains to the LLM for planning.
\end{itemize}

We present the experimental results in Table~\ref{tab:ablation_crag}. {Specifically, depth and breadth expansions are the most influential modules in CacheRAG: removing depth and breadth expansion leads to 22\% and 18.1\% performance drops in terms of truthfulness, while it results in a significant increase in the answering miss rate (26.1\% and 19\%). We attribute this phenomenon to the fact that removing depth and breadth expansions decreases the retrieval recall rate; as a result, the miss rate increases, and the overall accuracy and truthfulness decrease.} We further notice that example caching is another crucial component that significantly influences the performance of CacheRAG: by removing the hierarchical semantic cache, the accuracy and truthfulness drop by 4.3\% and 5.5\%, while the miss rate increases by 3\%, which indicates that the quality of generated retrieval plans declines when we remove example caching for the LLM planner to learn. The MMR, layered index, and auto-generation mechanism are also influential factors: removing these leads to truthfulness drops of 2.9\%, 2.8\%, and 1.8\%.

\begin{table}[hbtp]
  \caption{{Ablation Study.}}
  \small
  \vspace{-1em}
  \label{tab:ablation_crag}
\centering
\begin{tabular}{lrrrrr}
\toprule 
 \textbf{Model} & \textbf{Accu.} & \textbf{Hall.} & \textbf{Miss.} & \textbf{Truth.} \\
\midrule
CacheRAG w/o depth & 0.583 & \textbf{0.092} & 0.325 & 0.491\\
 CacheRAG w/o breadth & 0.638 & 0.108 & 0.254 & 0.530 \\
  CacheRAG w/o caching & 0.781 & 0.125 & 0.094 & 0.656  \\
  CacheRAG w/o MMR & 0.804 & 0.123 & 0.073 & 0.682 \\
  CacheRAG one layer & 0.808 & 0.125 & 0.067 & 0.683 \\
   CacheRAG w/o auto-gen & 0.811 & 0.118 & 0.071 & 0.693 \\
 %CacheRAG w/o domain & 0.815 & 0.131 & \textbf{0.054} & 0.684 \\
\cmidrule{1-5}
 CacheRAG   &     \textbf{0.824}               &                0.112        & \textbf{0.064}                  & \textbf{0.711}                            \\
\bottomrule
\end{tabular}
\vspace{-1em}
\end{table}

\subsection{LLM Model Robustness}
\label{subsec_robust}
{We evaluated the robustness of CacheRAG with different LLM backbones by replacing the LLM base model with three widely used LLMs: Llama-3.1-70B-Instruct, GPT-4o, and Claude-3.5-Sonnet and present the experimental results in Table~\ref{tab:backbone}. We observe that the performance of CacheRAG is relatively stable: it changes by up to 5\% when we replace the base model with other LLMs, which validates the robustness of the CacheRAG pipeline across different LLM backbones. }

\begin{table}[hbtp]
  \caption{{Robustness of LLM Backbones.}}
  \small
  \vspace{-1em}
  \label{tab:backbone}
\centering
\begin{tabular}{llrrrrr}
\toprule 
 \textbf{Model} & \textbf{Accu.} & \textbf{Hall.} & \textbf{Miss.} & \textbf{Truth.} \\
\midrule
  CacheRAG (Llama)  &    0.791                &        0.125              &    0.084            &  0.666            \\
   CacheRAG (GPT)   &  0.818 &  0.114 &  0.067 & 0.704 \\
    CacheRAG (Claude)   &  0.830 & 0.111 & 0.059 & 0.720 \\   
  CacheRAG (DeepSeek)  &     0.824          &       0.112             &  0.064                &     0.711         \\
\bottomrule
\end{tabular}
\end{table}

\subsection{Indexing Consistency and Robustness}
\label{subsec_consistency}
To validate the stability of our two-layer hierarchical index (Domain $\rightarrow$ Aspect) proposed in Section~\ref{subsec:diversity_cache}, we evaluate the consistency of the LLM-based logical parser. Inconsistencies in extracting the domain and aspect attributes could potentially lead to cache retrieval failures. To quantify this, we repeated the indexing process for each question in the CRAG dataset five times with the LLM temperature strictly set to 0. The empirical results show an extremely low inconsistent indexing rate of only $1.1\%$, confirming that the logical parsing phase is highly stable. Furthermore, CacheRAG is designed to be robust against such rare anomalies: if a user's question is assigned to a non-existent or empty aspect bucket, our retrieval algorithm (Algorithm~\ref{algo:cache}) deterministically relaxes the search space to the broader domain level. This fallback mechanism ensures that relevant historical plans can always be retrieved, effectively mitigating the impact of any parsing inconsistencies.

\subsection{Efficiency}
\label{subsec_efficiency}

We report the mean inference time of CacheRAG against the baselines on CRAG in this section. Specifically, the mean inference time of CacheRAG is 9.44s {(Logical Parsing and KG Routing, Hierarchical Cache Retrieval, Physical Plan Compilation, Bounded Depth and Breadth Expansion, Heuristic Dispatching, and Summarization take 2.75s, 0.02s, 0.49s, 3.93s, 0.47s, and 1.78s)}, while the state-of-the-art baseline, Apex, takes 5.96s. Although the inference time of CacheRAG is higher than that of the state-of-the-art solution on CRAG, we believe this is reasonable, as our approach aims to improve the effectiveness of existing methods and additionally requires depth and breadth expansion of KG paths, which needs more time to complete. In other words, our approach trades efficiency for effectiveness. However, we would like to point out that the mean inference times of CacheRAG and the state-of-the-art solution, Apex, are of the same order of magnitude. Therefore, we believe the inference time of our proposed CacheRAG approach is generally acceptable.

\begin{figure}[hbtp]
  \centering
\includegraphics[width=\linewidth]{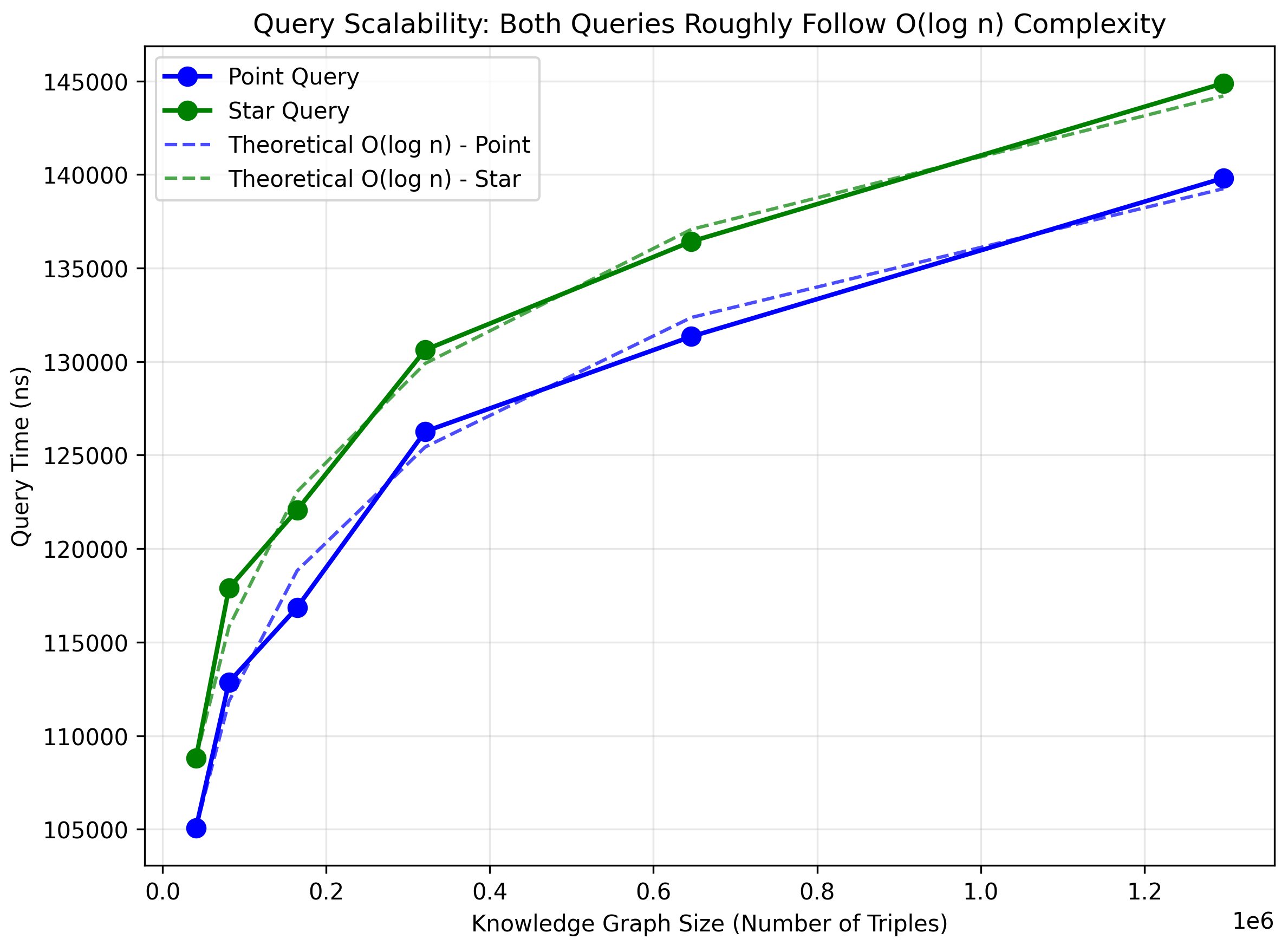}
\vspace{-1em}
      \caption{Scalability Experiments.}
  \label{fig:scalability}
  \vspace{-1em}
\end{figure}

\subsection{Time and Memory Complexity and Scalability}
\label{subsec_scalability}
We analyze the time and memory complexity of CacheRAG using the formal bounds defined in Section~\ref{sec_method}. For semantic caching, the hierarchical index routing takes $\mathcal{O}(1)$ dictionary lookup time, while the MMR scheduling takes $\mathcal{O}(b \log b)$ time, utilizing $\mathcal{O}(N)$ total space (where $b \ll N$ is the localized bucket size and $N$ is the global cache size). During execution, the Bounded Depth Expansion (Index Nested Loop Join) takes $\mathcal{O}(K_{depth} \log n)$ time and $\mathcal{O}(K_{depth})$ space, where $K_{depth} \le 3$ is the maximum hop limit and $n$ is the total number of KG triples. The Bounded Breadth Expansion (Star-Pattern Scan) processes the frontier nodes, taking $\mathcal{O}(K_{depth} \cdot (K_{degree} + \log n))$ time and strictly $\mathcal{O}(K_{depth} \cdot K_{degree})$ space, where $K_{degree}$ is the Top-$K$ pruning bound. The total time and memory induced by LLM interactions (Parser, Compiler, Dispatcher, and Summarizer) are $\mathcal{O}(L \cdot K_{depth})$ and $\mathcal{O}(K_{depth})$, where $L$ is the bounded LLM inference latency. The overall online time complexity is $\mathcal{O}(b \log b + K_{depth} \log n + K_{depth} \cdot K_{degree} + L \cdot K_{depth})$. Since $b$, $K_{depth}$, $K_{degree}$, and $L$ are treated as strict system constants, CacheRAG is highly scalable. Detailed step-by-step derivations are presented in Appendix~\ref{sec_appendix_scalability}.

\paragraph{Empirical Scalability on Synthetic KGs.} To empirically validate the theoretical complexity, we evaluated the retrieval latency on synthetic KGs of varying sizes. Following the schema of Wikidata, we generated synthetic KGs containing from 40,000 up to 1.28 million triples, with entity degrees approximating Wikidata's actual distribution. We executed 50 point queries (depth expansion) and 50 star queries (breadth expansion) across these KGs. As shown in Figure~\ref{fig:scalability}, the execution time for both query types exhibits a clear logarithmic growth $\mathcal{O}(\log n)$ as the KG size increases. This confirms that CacheRAG's bounded graph operators are highly scalable and the retrieval mechanism will not become a performance bottleneck on larger-scale graphs.

\subsection{LRU Cache Design}
\label{subsec_lru}
{To further explore the design of two-layered indexing, we implemented a Least Recently Used (LRU) CacheRAG variant (CacheRAG (LRU)), where a maximum of 10 examples are kept under each bucket. This design is suitable for scenarios with limited caching storage space. Compared to storing all historical data, we saved $64\%$ of space on CRAG dataset. As shown in Table~\ref{tab:lru}, the performance drops by less than $2\%$ after implementing the LRU caching, validating the robustness of the LRU-based caching variant of CacheRAG.}

\begin{table}[hbtp]
  \caption{{LRU-based variant results on CRAG.}}
  \small
  \vspace{-1em}
  \label{tab:lru}
\centering
\begin{tabular}{llrrrrr}
\toprule 
 \textbf{Model} & \textbf{Accu.} & \textbf{Hall.} & \textbf{Miss.} & \textbf{Truth.} \\
\midrule
 CacheRAG (LRU)   &  0.808   & 0.116    &  0.076   &  0.692   \\
  CacheRAG (DeepSeek)  &     0.824          &       0.112             &  0.064                &     0.711         \\
\bottomrule
\end{tabular}
\end{table}

%% file: secs/sec_conclusion.tex
\section{Conclusion}
\label{sec:conclusion}

This paper presents CacheRAG, a systematic architecture that transforms stateless LLM planners into continual learners for Knowledge Graph Question Answering. By synergistically integrating a schema-constrained semantic parser, a diversity-aware hierarchical cache, and deterministically bounded graph operators, CacheRAG resolves critical execution bottlenecks like schema hallucination and combinatorial explosion. Extensive evaluations validate this stateful paradigm: CacheRAG significantly outperforms state-of-the-art baselines, achieving a 13.2\% accuracy and 17.5\% truthfulness increase on the complex CRAG benchmark. Furthermore, it generalizes robustly to SPARQL-based KGs and scales logarithmically. Ultimately, CacheRAG provides a principled foundation for embedding history-aware, verifiable continual learning into future LLM-augmented database systems.

%% file: secs/appendix.tex
\section{Appendix}

\subsection{Dataset Details}
\label{subsec_datasets_details}
To evaluate the performance of our method, we consider the widely used CRAG~\cite{yang2024crag} benchmark, where the KGs cover five distinct domains and are accessed through API functions. We use the KG questions in CRAG to conduct experiments.

The statistics of the CRAG KG private testing set are presented in Table~\ref{tab:crag-stats}. CRAG categorizes the questions into head, torso, and tail based on the popularity of the topic entity. It also considers the domain, question types, and timeliness of the questions, categorizing them into different subsets. In total, we experimented on the head, torso, and tail question splits, which have 187, 203, and 188 questions, respectively.

% music 7 31 7
\begin{table}[hbtp]
\centering
\caption{Statistics of each question type in different question splits of CRAG.}
\label{tab:crag-stats}
\begin{tabular}{llll}
\toprule
                     & \textbf{head} & \textbf{torso} & \textbf{tail} \\
\midrule
open                 &  24  & 24   & 24   \\
finance              & 57  &  59  &  58 \\
movie                & 60  & 58   & 59  \\
music                &  7  &  31   &   7 \\
sports               &  39  &  31   &  40  \\
\midrule
simple               & 98  &  104  &  94 \\
simple\_w\_condition & 23   &  26   & 28   \\
set                  &  9  &  6   & 5   \\
comparison           &  17  & 22    & 21   \\
aggregation          & 16   & 18    & 21   \\
multi-hop            &  8  &  9   &   6 \\
post-processing      &  3   &   3   &   3  \\
false premise        &  13  & 15    & 10   \\
\midrule
real-time            & 32   &  38   &  36  \\
fast-changing        &  31  &  32   &   39 \\
slow-changing        &  22  &  27   &  19  \\
static               & 102  &  106  &  94 \\
\midrule
total               & 187 & 203 & 188 \\
\bottomrule
\end{tabular}
\end{table}

We also included SPARQL-based datasets used by state-of-the-art approaches: QALD-10-en~\cite{Usbeck2023QALD10T}, WebQSP~\cite{yih2016value}, and CWQ~\cite{talmor-berant-2018-web} datasets to comprehensively evaluate the performance of CacheRAG.

\subsection{Metric and Evaluation Scheme Details}
\label{subsec_metric_details}

We present the details on the reliability of the auto evaluation critic in Table~\ref{tab:auto-eval}.

\begin{table}[htbp]
\caption{The Auto-evaluation Accuracy of Llama-3.1 70B Instruct Model.}
\label{tab:auto-eval}
\centering
\begin{tabular}{ll}
\toprule
                    & \textbf{Llama-3.1 70B} \\
\midrule
\textbf{correct answers}     & 98.2\%        \\
\textbf{incorrect answers}   & 96.8\%        \\
\textbf{missing}             & 100\%         \\
\textbf{overall reliability} & 98.4\%       \\
\bottomrule
\end{tabular}
\end{table}

\subsection{Baseline Details}
\label{subsec_baseline_details}

We compare our approach against the following baselines, which can be categorized into LLM base models, LLM tool-calling models, KDD Cup-winning solutions, and SPARQL-based state-of-the-art solutions.

\subsubsection{LLM Base Models}
On CRAG dataset, we include \textit{GPT-4o~\cite{hurst2024gpt}}, \textit{Llama-3.1-70B-Instruct~\cite{dubey2024llama}}, and \textit{Deepseek-chat-V3-0324~\cite{liu2024deepseek}} base models as baselines, which covers the state-of-the-art open-sourced and close-sourced LLMs.

\subsubsection{LLM Tool Calling Models}
On CRAG, we further include \textit{StructGPT~\cite{jiang2023structgpt}}, which enables GPT-4o, Llama-3.1-70B-Instruct, and Deepseek-chat-V3-0324 base models to freely perform tool calling. Specifically, we provide the metadata of the API functions (\textit{i.e.}, the function name, parameters, descriptions, and sample use cases) so that the LLMs can freely chain and call the API functions to retrieve relevant content from the KGs based on their own planning and reasoning abilities.

\subsubsection{KDD Cup Winning Solutions}
Since CRAG is used to host the KDD Cup competition in 2024, we also include the winning solutions to establish the state-of-the-art baselines on CRAG:
\begin{itemize}
	\item \textit{db3~\cite{xia2024winning}:} The design of db3 jointly considers inputs from the KG and web content. We enter ``<EMPTY>'' into the web content module of db3 to adapt it to our KGQA settings. A router-based adaptive RAG pipeline is introduced in apex to retrieve relevant KG content for answer generation. We use the Deepseek-chat-V3-0324 model as the base model.
	\item \textit{apex~\cite{ouyang2024revisiting}:} A router-based adaptive RAG pipeline is introduced to retrieve relevant KG content for answer generation. Similar to db3, we input ``<EMPTY>'' as the web content input to the model to adapt the apex solution to our KGQA settings. We use the Deepseek-chat-V3-0324 model as the base model.
\end{itemize}

\begin{figure*}[hbtp]
\centering
\begin{subfigure}{0.23\linewidth}
  \includegraphics[width=\linewidth]{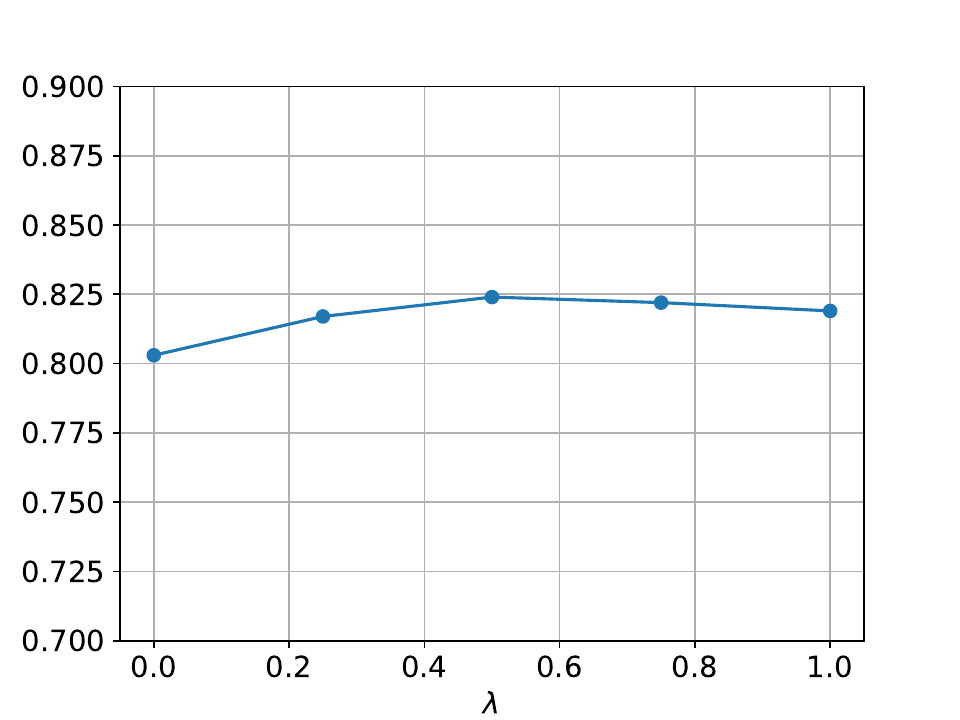}
  \caption{accuracy}
  \label{subfig:acc}
\end{subfigure}
\begin{subfigure}{0.23\linewidth}
  \includegraphics[width=\linewidth]{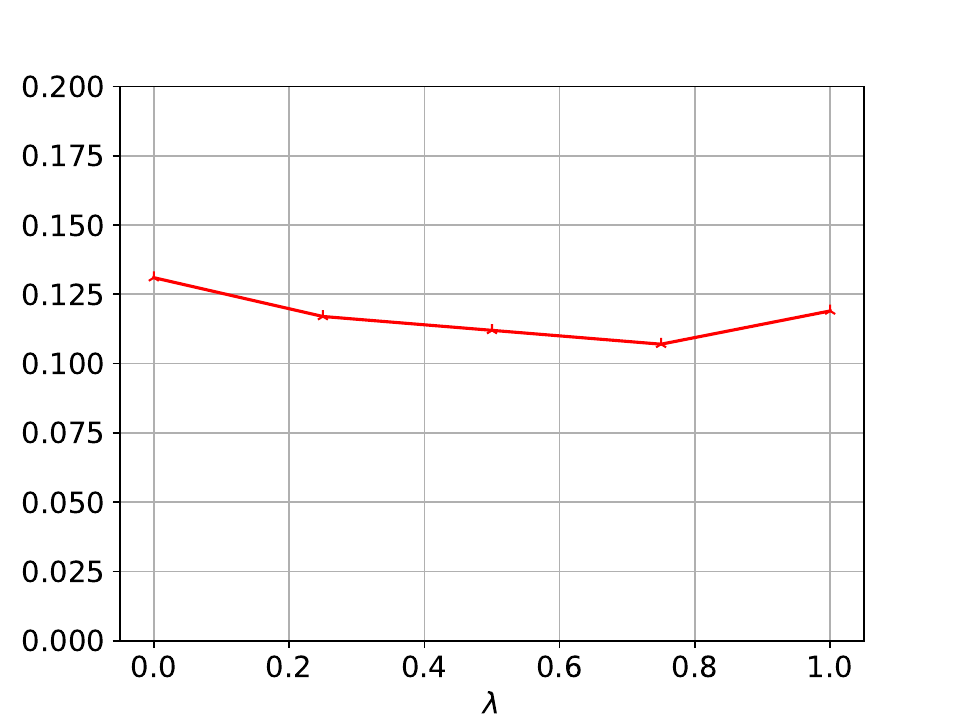}
  \caption{hallucination}
  \label{subfig:hall}
\end{subfigure}
\begin{subfigure}{0.23\linewidth}
  \includegraphics[width=\linewidth]{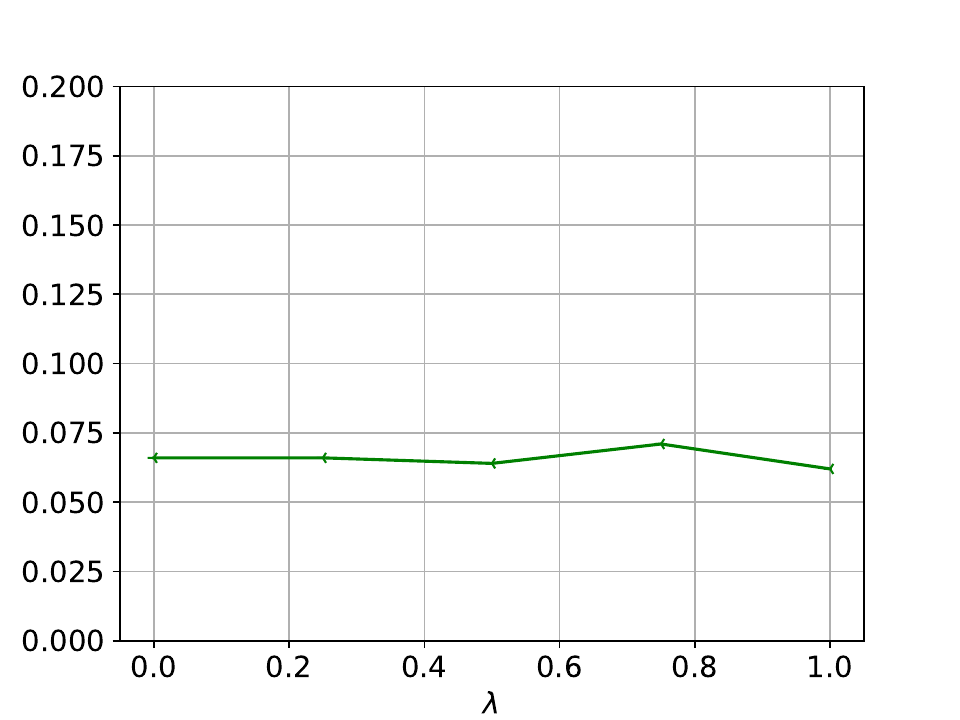}
  \caption{miss}
  \label{subfig:miss}
\end{subfigure}
\begin{subfigure}{0.23\linewidth}
  \includegraphics[width=\linewidth]{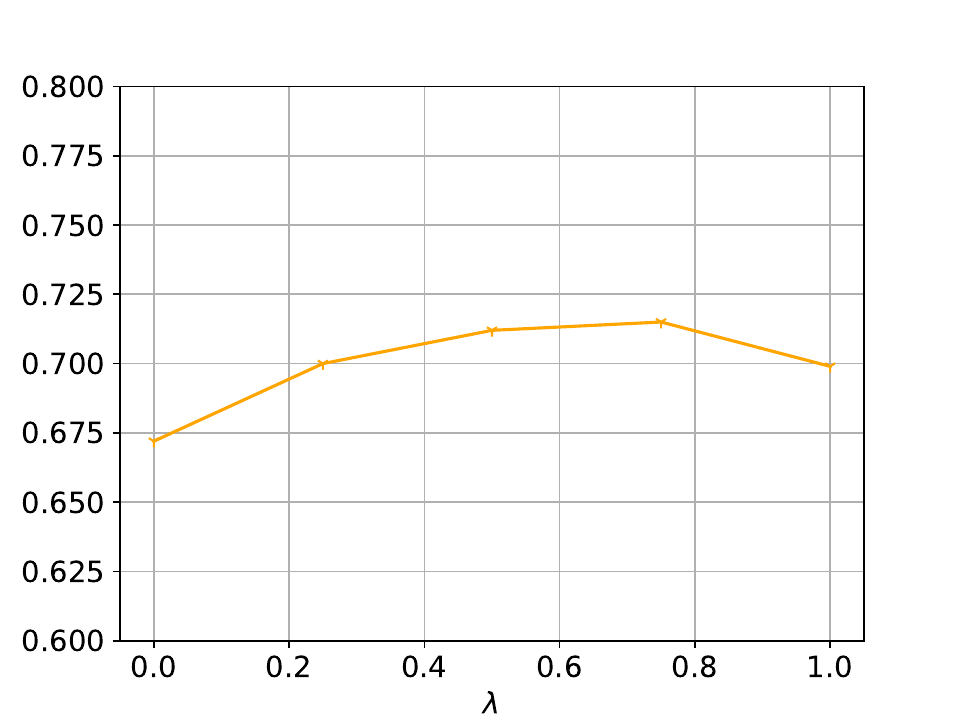}
  \caption{truthfulness}
  \label{subfig:truth}
\end{subfigure}
  \caption{The parameter $\lambda$'s sensitivity of CacheRAG on CRAG dataset.}
  \label{fig:para}
\end{figure*}

\subsubsection{SPARQL-based State-of-the-art Solutions}
On the SPARQL-based datasets QALD-10-en, WebQSP, and CWQ, we consider the following state-of-the-art methods:
\begin{itemize}
	\item \textit{ToG~\cite{sunthink}:} ToG integrates large language models (LLMs) with knowledge graphs (KGs). Through beam search, LLMs are used to iteratively explore reasoning paths on KGs to enhance the deep reasoning ability of LLMs, improve knowledge traceability and correctness, and achieve state-of-the-art performance on multiple datasets.
	\item \textit{ToG-2~\cite{ma2025thinkongraph}:} Following ToG, ToG-2 is a hybrid RAG framework that tightly couples knowledge graphs and documents for iterative retrieval, enabling deep and faithful reasoning in LLMs with state-of-the-art performance on multiple datasets. Apart from the KG itself, ToG-2 additionally relies on external Wikipedia pages as an auxiliary knowledge source.
	\item \textit{sparql-qa~\cite{borroto2022sparql}:} sparql-qa uses a neural architecture combining NMT and NER with input processing and QQT format to translate natural language into SPARQL.
	\item \textit{Decaf~\cite{yu2022decaf}:} Decaf jointly generates answers and logical forms for knowledge base question answering, combines the advantages of both forms, and uses text retrieval instead of entity linking to enhance generality.
	\item \textit{CBR~\cite{das2021case}:}  CBR is a neuro-symbolic method. It retrieves similar question cases, reuses their logical form components, and revises the generated form using KB embeddings to handle complex KBQA and unseen relations.
\end{itemize}

Note that since the API-based CRAG dataset does not support SPARQL querying, so we ran them only on SPARQL-based datasets.

\subsection{Parameter Sensitivity}
\label{subsec_para}
In this section, we analyze the sensitivity of an important hyperparameter: the retrieval balancing parameter $\lambda$. Specifically, we experimented with CacheRAG by setting $\lambda$ to 0, 0.25, 0.5, 0.75, and 1, and we present the corresponding accuracy, hallucination rate, miss rate, and truthfulness score in Figure~\ref{fig:para}. As shown in Figure~\ref{fig:para}, CacheRAG achieves relatively stable performance when $\lambda \in [0.25, 1]$ (up to $0.7\%$ change in terms of accuracy and up to $1.5\%$ change in terms of truthfulness). The model achieves the highest accuracy when $\lambda=0.5$, while it achieves the lowest hallucination rate when $\lambda=0.75$. In general, we believe selecting $\lambda=0.5$ could be an intuitive and effective choice that balances the diversity and relevance of learning samples for the LLM planner.

\subsection{Scalability Details}
\label{sec_appendix_scalability}

We analyze the scalability of CacheRAG step by step, utilizing the formal architectural bounds established in our methodology:

1) \textbf{Diversity-Aware Cache}: The caching mechanism's efficiency is decoupled from the KG scale. Navigating the two-layer hierarchical index (Domain $\rightarrow$ Aspect) requires $\mathcal{O}(1)$ dictionary lookup time. Evaluating structural diversity via MMR within the localized bucket takes $\mathcal{O}(b \log b)$ time, where $b$ is the bounded bucket capacity. The global space complexity is $\mathcal{O}(N)$, where $N$ is the total number of cached historical plans.

2) \textbf{Bounded Depth Expansion ($\sigma_{depth}$)}: As an Index Nested Loop Join, the queries generated in this step are point queries:
\begin{verbatim}
SELECT ?value WHERE {
:specific_entity :specific_attribute ?value .
}
\end{verbatim}
Based on the SPO B+ Tree indices used by KGs like Wikidata~\cite{blazegraph-wikidata, fahl2022getting}, a single index lookup induces $\mathcal{O}(\log n)$ time complexity, where $n$ is the total number of KG triples. Since this operator is strictly bounded to trigger at most $K_{depth}$ times, the total time complexity is $\mathcal{O}(K_{depth} \log n)$. Retaining the intermediate trajectory states requires $\mathcal{O}(K_{depth})$ memory.

3) \textbf{Bounded Breadth Expansion ($\sigma_{breadth}$)}: This operator executes bounded Star-Pattern Neighborhood Scans:
\begin{verbatim}
SELECT ?property ?value WHERE {
:specific_entity ?property ?value .
}
\end{verbatim}
Fetching the subject via the B+ Tree takes $\mathcal{O}(\log n)$ time, and sequentially scanning to retrieve up to $K_{degree}$ top-ranked neighbors takes $\mathcal{O}(K_{degree})$ time, yielding $\mathcal{O}(K_{degree} + \log n)$ per entity. Applied across the frontier nodes (bounded by $K_{depth}$), the total time complexity is $\mathcal{O}(K_{depth} \cdot (K_{degree} + \log n))$. Retaining this local topology strictly requires $\mathcal{O}(K_{depth} \cdot K_{degree})$ memory.

4) \textbf{LLM Execution Overhead}: The number of API invocations to the LLM is deterministic: 1 call for the \textbf{Logical Parser}, 1 for the \textbf{Backend Adapter}, up to $K_{depth}$ loop evaluations for the \textbf{Heuristic Dispatcher}, and 1 final call for the \textbf{Summarizer}. Therefore, the total number of LLM inferences is bounded by $\mathcal{O}(K_{depth})$. Assuming a maximum LLM inference latency $L$, the time complexity induced by the LLM is $\mathcal{O}(L \cdot K_{depth})$. Storing the intermediate logical contexts takes $\mathcal{O}(K_{depth})$ memory.

\textbf{Overall Complexity}: $K_{depth}$ and $K_{degree}$ are administrative system bounds designed to prevent combinatorial explosion (e.g., $K_{depth} \le 3$, $K_{degree} \le 30$). The bucket size $b$ and LLM latency $L$ are similarly treated as system constants. Therefore, the overall online time complexity evaluates to $\mathcal{O}(b \log b + K_{depth} \log n + K_{depth} \cdot K_{degree} + L \cdot K_{depth}) = \mathcal{O}(\log n + L)$, and the space complexity evaluates to $\mathcal{O}(N + K_{depth} + K_{depth} \cdot K_{degree}) = \mathcal{O}(N + K_{depth} \cdot K_{degree})$. The critical finding is that the online execution time grows strictly logarithmically ($\mathcal{O}(\log n)$) with respect to the underlying knowledge graph size, demonstrating excellent production scalability.

\subsection{Head Torso Tail Analysis}
\label{subsec_htt}

To comprehensively study the performance of CacheRAG in head, torso, and tail data splits, we evaluate the performance of CacheRAG across the three CRAG data splits and compare it with the base LLM in Table~\ref{tab:baseline_benchmarking_crag_full}. Specifically, the base LLMs GPT-4o, Llama-3.1-70B-Instruct, and Deepseek-chat-V3-0324 show a significant decreasing trend in terms of accuracy and truthfulness scores as we move from head to tail data splits (up to 14\%, 17\%, and 17\% for the three models, respectively). Our CacheRAG model demonstrates relatively stable performance across head, torso, and tail (with a 4.2\% decrease in accuracy and a 3.6\% decrease in truthfulness score), which illustrates the robustness of our method concerning the varying popularity of the questions.

\begin{table}[hbtp]
  \caption{Head/torso/tail experimental results on CRAG.}
  \label{tab:baseline_benchmarking_crag_full}
\centering
\begin{tabular}{llrrrrr}
\toprule &
 \textbf{Model} & \textbf{A} & \textbf{H} & \textbf{M} & \textbf{T} \\
\midrule
head & GPT-4o   &     0.401               &             \textbf{0.086}           &    0.513               &      0.316          \\
& Llama   &   0.390                 &         0.096               &        0.513           &    0.294            \\
& Deepseek & 0.487 & 0.134 & 0.380 & 0.353 \\
\cmidrule{2-6}
& CacheRAG   &     \textbf{0.840}               &  0.123                      &   \textbf{0.037}                &     \textbf{0.717}           \\
\midrule
torso & GPT-4o   &      0.355              &                 \textbf{0.089}       &    0.557               &       0.266         \\
& Llama   &      0.305              &     \textbf{0.089}                   &        0.606           &   0.217             \\
& Deepseek & 0.394 & 0.158 & 0.448 & 0.236 \\
\cmidrule{2-6}
& CacheRAG   &     \textbf{0.833}               &     0.098                   &      \textbf{0.069}             &     \textbf{0.735}          \\
\midrule
tail & GPT-4o   &        0.266            &    0.096                    &    0.638               &       0.170        \\
& Llama   &    0.223                &          \textbf{0.053}              &       0.723            &   0.170             \\
& Deepseek & 0.319 & 0.128 & 0.553 & 0.191 \\
\cmidrule{2-6}
& CacheRAG   &     \textbf{0.798}            &    0.117                    &   \textbf{0.085}              &  \textbf{0.681}             \\
\bottomrule
\end{tabular}
\end{table}

\subsection{Multi-KG Routing}
\label{subsec_multi}

As mentioned in Section~\ref{subsec_datasets}, the CRAG dataset contains five KGs from five different domains. If the LLM planner is to correctly apply the API functions, it needs to be routed to the correct KG. Therefore, we would like to analyze CacheRAG's performance on multi-KG domain routing. Specifically, we compare the performance of the following domain routing designs:

\begin{itemize}
    \item \textit{a) Naive Llama-3.1-8B-Instruct model: }Directly prompt the LLM with the question and KG names to know which domain's KG can solve the user's question.
    \item \textit{b) Llama-3.1-8B-Instruct model + static KG descriptions: }Prompt the LLM with the question and static brief KG descriptions (The ``Initial KG descriptions'' column of Table~\ref{tab:example-domain}) to route the user questions.
    \item \textit{c) Llama-3.1-8B-Instruct model + static KG descriptions + Chain-of-Thoughts reasoning: }Prompt the LLM with the question and static brief KG descriptions in a Chain-of-Thoughts manner to route the user questions.
    \item \textit{d) Our design (Llama-3.1-8B-Instruct model + dynamic KG descriptions + Chain-of-Thought reasoning): }Based on design c), we prompt the LLM to maintain a dynamic KG description cache. Specifically, if the domain is incorrectly identified, we prompt the LLM to update the KG descriptions so that the LLM-based domain router can better understand the KGs by referencing the KG descriptions generated by itself.
\end{itemize}

To further explain how the dynamic KG description works, we present the Multi-KG Domain Description Update template. 

\begin{block}{Multi-KG Domain Description Update Skeleton}
You will be given set of KGs with descrptions, your task is to refine the descriptions of the KGs.

These are the current descriptions of the KGs:

[$K_i$]: [$K_i$'s descriptions]

...

In the previous round, you were asked to determine which external data source is most suitable for answering this question. [Task descriptions].

Based on the KG descriptions, you thought the most appropriate data source was [Previous Domain Selection], for the question [$Q_i$].

Your reasoning process was [Previous Reasoning Steps].

Your selection was wrong.

Based on this answering history, please update the KG descriptions to better answer the question in the next round. Please focus on identifying the differences between different KGs, try to avoid overlapping descriptions for different KGs! 

[Further Instructions] (Optional)

\end{block}

\begin{figure}[hbtp]
  \centering
\includegraphics[width=0.9\linewidth]{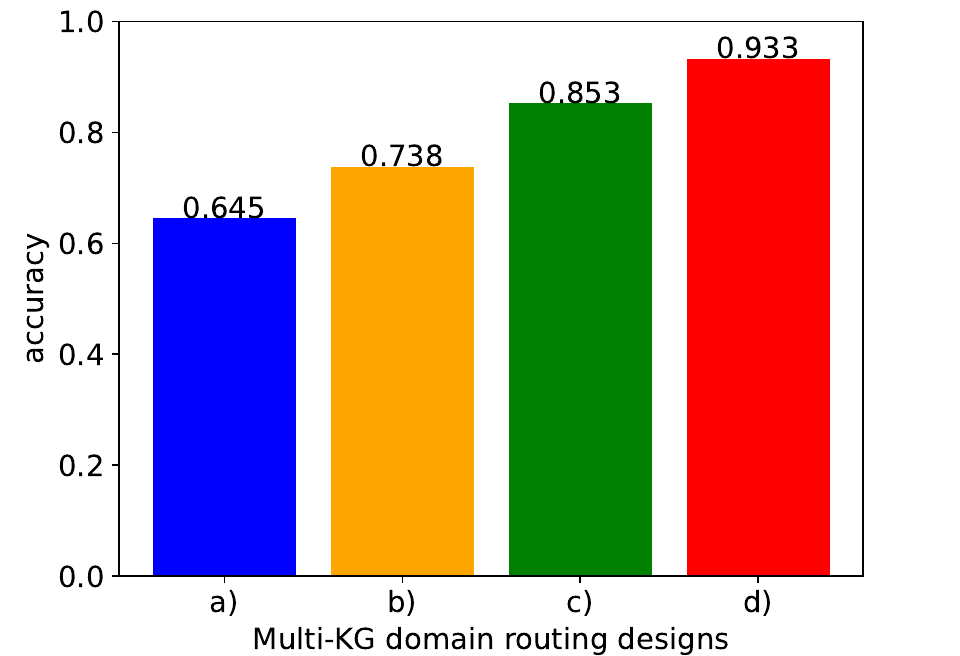}
      \caption{The multi-KG domain routing accuracy of different designs. Our Llama-3.1-8B-Instruct + dynamic KB descriptions + Chain-of-Thoughts reasoning design outperforms the rest of the domain routing design by $8\%$.}
  \label{fig:domain-routing}
\end{figure}

We also present the initial KG descriptions and the updated final KG descriptions in Table~\ref{tab:example-domain}. As shown in Table~\ref{tab:example-domain}, the initial KG descriptions are enriched by the LLM router to be more precise and accurate in understanding the content, coverage, and functionality of each respective KG. We present the quantitative experimental results in Figure~\ref{fig:domain-routing}. In general, our design outperforms other multi-KG domain routing designs by $8\%$, demonstrating the effectiveness of CacheRAG's dynamic KG description cache.

\begin{table*}[hbtp]
	
	\centering
	\caption{Exemplar initial KG descriptions and the updated final KG descriptions.}
	\label{tab:example-domain}
	\begin{tabular}{p{1cm} p{8cm} p{8cm}}
		\toprule
		\textbf{Domain}             & \textbf{Initial KG descriptions} & \textbf{Final KG descriptions} \\    
		\midrule
		Open & This KG includes content in Open domain. The content is based on Wikidata, you can use it as a general encyclopedia. & This KG includes general knowledge about the world, excluding specific domains like film, finance, and sports, but including information about history, science, culture, and general information about people, organizations, and events that are not related to any specific domain.\\
            Movie & This KG includes content in Movie domain. The content is based on IMDB, you can find the detailed information of the actors, movies, and oscar awards. & This KG is focused on the film industry, including information about movie titles, film scripts, and film-related events, but excluding information about music releases, financial transactions, general knowledge about the world, and sports.\\
            Music & This KG includes content in Music domain. The content is based on musicBuzz and Billboard, you can find the detailed information of the singers, albums, songs, and billboard results. & This KG is focused on the music industry, including information about musicians, music genres, album releases, and music-related events, but excluding information about film releases, financial transactions, general knowledge about the world, and sports. \\
            Finance & This KG includes content in Finance domain. The content is based on Yahoo finance, you can find the detailed information of the stock prices, eps, p/e ratio, etc. & This KG is focused on financial transactions and market trends, including economic indicators, stock market data, and financial news, with a focus on specific financial data, such as stock prices, financial reports, and market analysis.\\
            Sports & This KG includes content in Sports domain. The content is based on basketball and soccer, you can find the detailed information of the NBA and Premier League match results and team leaders. & This KG is focused on the competitive aspects of sports, including information about teams, players, championships, and tournaments, as well as the rules, strategies, and techniques of specific sports.\\
		\bottomrule
	\end{tabular}
\end{table*}